\documentclass[10pt,preprint]{aastex}
\usepackage{color}

\begin{document}

\title{Pulsational Analysis of the Cores of Massive Stars and its
  Relevance to Pulsar Kicks}

\author{Jeremiah W. Murphy\altaffilmark{1,2}}
\author{Adam Burrows\altaffilmark{1}}

\and

\author{Alexander Heger\altaffilmark{3,4}}

\altaffiltext{1}{Steward Observatory, The University of Arizona,
  Tucson, AZ 85721; jmurphy@as.arizona.edu, aburrows@as.arizona.edu}
\altaffiltext{2}{JINA Fellow.}
\altaffiltext{3}{Theoretical Astrophysics Group, T-6, MS B227, Los
  Alamos National Laboratory, Los Alamos, NM 87544, U.S.A.; alex@t6.lanl.gov }
\altaffiltext{4}{Enrico Fermi Institute, The University of Chicago, 5640 S Ellis Ave., Chicago, IL, 60637, U.S.A.}

\begin{abstract}
The mechanism responsible for the natal kicks of neutron stars
continues to be a challenging problem.  Indeed, many mechanisms have
been suggested, and one hydrodynamic mechanism may
require large initial asymmetries in  the cores of supernova progenitor stars.
\markcite{goldreich97}{Goldreich}, {Lai}, \&  {Sahrling} (1997) suggested that unstable g-modes
trapped in the iron (Fe) core by the convective burning layers and excited by
the $\epsilon$-mechanism may provide the requisite asymmetries.  We
perform a modal analysis of the last minutes before collapse of
published core structures and derive eigenfrequencies and eigenfunctions, including the nonadiabatic
effects of growth by nuclear burning and decay by both neutrino and
acoustic losses.  In general, we find two types of g-modes:
inner-core g-modes, which are stabilized by neutrino losses and
outer-core g-modes which are trapped near the burning shells and can
be unstable.  Without exception, we find at least one unstable g-mode
for each progenitor in the entire mass range we consider, 11 M$_{\sun}$ to
40 M$_{\sun}$.  More importantly, we find that the timescales for
growth and decay are an order of magnitude or more longer than the
time until the commencement of core collapse.  We conclude that the
$\epsilon$-mechanism may not have enough time to significantly amplify
core g-modes prior to collapse.
\end{abstract}


\section{Introduction}
\label{sec:intro}
Observations indicate that many pulsars have high proper motions
\markcite{lyne94,lorimer97,hansen97,cordes98}({Lyne} \& {Lorimer} 1994; {Lorimer}, {Bailes}, \&  {Harrison} 1997; {Hansen} \& {Phinney} 1997; {Cordes} \& {Chernoff} 1998) and that many neutron star
 systems require kicks at birth.  For example, pulsar bow shock morphologies
\markcite{cordes93}({Cordes}, {Romani}, \& {Lundgren} 1993), misaligned spins of binary pulsars
\markcite{cordes90,kramer98,wex00,kaspi96,lai95,lai96,kumar97}({Cordes}, {Wasserman}, \&  {Blaskiewicz} 1990; {Kramer} 1998; {Wex}, {Kalogera}, \& {Kramer} 2000; {Kaspi} {et~al.} 1996; {Lai}, {Bildsten}, \& {Kaspi} 1995; {Lai} 1996; {Kumar} \& {Quataert} 1997), the
statistics of low-mass X-ray binaries (LMXBs)
\markcite{kalogera97,kalogera98}({Kalogera} 1997; {Kalogera} \& {Webbink} 1998), the high orbital eccentricities of some
Be star systems \markcite{verbunt95}({Verbunt} \& {van den Heuvel} 1995), and scenarios for double neutron star systems
\markcite{dewey87,fryer97,fryer98,dewi04,willems04}({Dewey} \& {Cordes} 1987; {Fryer} \& {Kalogera} 1997; {Fryer}, {Burrows}, \& {Benz} 1998; {Dewi} \& {van den Heuvel} 2004; {Willems} \& {Kalogera} 2004) all indicate that neutron stars
experience significant natal kicks. Furthermore, \markcite{arzoumanian02}{Arzoumanian}, {Chernoff}, \&  {Cordes} (2002)
find a bimodal distribution of birth kick speeds with peaks at $\sim$100
km s$^{-1}$ and $\sim$700 km s$^{-1}$ and dispersions of 90 and 500
km s$^{-1}$.  They suggest that roughly 50\% of the pulsars in
the solar neigborhood escape the Galaxy and that $\sim$15\% have
kick speeds greater than 1000 km s$^{-1}$.  The observations indicate that kicks seem to be hallmarks of neutron star formation.

What is the mechanism producing the observed and inferred kicks?  To
give a $1.4$
M$_{\sun}$  neutron star a speed of 1000 km s$^{-1}$, the mechanism must be
able to impart $\sim$10$^{49}$ erg of kinetic energy, which is just
1\% of the canonical SN mechanical energy ($\sim$10$^{51}$ erg) or
0.01\% of the collapse energy ($\sim$10$^{53}$ erg).  Additionally, any
explanation must address the distribution of speeds.  If the
distribution is bimodal, then it is possible that more than one
mechanism is responsible for the kicks.  \markcite{lai00c}{Lai} (2000b) has summarized
the possible magnetic, neutrino ($\nu$), and hydrodynamic kick
mechanisms, but in general many of the scenarios
have trouble producing kicks larger than $\sim$200 km s$^{-1}$.

Of the many mechanisms propounded, hydrodynamic mechanisms
are the most compelling.  For example,
neutrino-driven convection, which is inherently a
nonspherical phenomenon, excites large-scale convective motions that
stochastically jostle the protoneutron star \markcite{burrows95,janka94}({Burrows}, {Hayes}, \& {Fryxell} 1995; {Janka} \& {M\"{u}ller} 1994).  However, multi-dimensional calculations indicate
that any kick imparted as a result of the convective jostling
(``Brownian motion'') has an
average magnitude of $\sim$200 km s$^{-1}$ \markcite{burrows95,janka94,scheck04}({Burrows} {et~al.} 1995; {Janka} \& {M\"{u}ller} 1994; {Scheck} {et~al.} 2004) and cannot explain the highest
  velocities nor the second peak in a possible bimodal distribution.  

These analyses suggest the need for additional asymmetries before the onset of collapse. \markcite{burrows96}{Burrows} \& {Hayes} (1996) state that global
asymmetries in the progenitor may be
prerequisites for large kicks.  In a 2-D
hydrodynamic simulation and with modest asymmetries,
\markcite{burrows96}{Burrows} \& {Hayes} (1996) are able to produce
a recoil speed of $\sim$550 km s$^{-1}$.  However, these calculations
are crude and the results, while suggestive, are not conclusive.  \markcite{goldreich97}{Goldreich} {et~al.} (1997) have proposed
that these asymmetries may be excited by the
$\epsilon$-mechanism in which g-mode
oscillations in the core are pumped by nuclear reactions.

As a prelude to more definitive multi-dimensional radiation
hydrodynamic investigations of neutron star kicks in the supernova context, one
would like to determine whether global asymmetries prior to core
collapse in fact occur.  Hence, a thorough analysis of the progenitor
oscillations and their growth rates is necessary. In this
paper, we explore the viability of the g-mode oscillation origin for
putative pre-collapse anisotropies by performing such an eigenmode
analysis.

\section{Linear Perturbation Equations}
To set the stage for our eigenmode analysis of the cores of
nonrotating massive
stars, we first describe linear stellar pulsation theory in the
general case.  In linear perturbation theory, a star is assumed
to be in hydrostatic equilibrium, around which perturbations are small,
making second-order or higher terms in the perturbation equations
negligibly small.  Therefore, if $f_0(\vec{r})$ is a quantity satisfying the
hydrostatic equations, then $f(\vec{r},t) \rightarrow f_0(r) +
f'(\vec{r},t)$, where $0$  denotes the background solution and $'$
denotes the Eulerian perturbation.  We assume that the perturbation
is of the form $f'(\vec{r},t) = f'(r)Y_{lm}(\theta,\phi)e^{i \omega
  t}$, where $Y_{\ell m}(\theta,\phi)$ is a spherical harmonic in that
$\ell$ and $m$ are the polar and azimuthal indeces, repsectively,  and $\omega$
is the oscillation frequency.  Under these assumptions and an assumption
of adiabatic perturbations, if $P$, $\rho$, $\phi$,
$c_s$ are the pressure, density, gravitational potential, and sound
speed, respectively, the adiabatic linear peturbation
equations\footnote{For a complete derivation of the
  adiabatic linear peturbation equations see
\markcite{unno89}{Unno} {et~al.} (1989).} are
\begin{equation}
\label{eq:nao1}
\frac{1}{r^{2}}\frac{d}{dr} (r^2\xi_r) - \frac{g}{c^2_s} \xi_r + \left
(1 - \frac{L^2_l}{\omega^2} \right ) \frac{P'}{\rho c_s^2} =
\frac{l(l+1)}{\omega^2 r^2} \phi',
\end{equation}
\begin{equation}
\label{eq:nao2}
\frac{1}{\rho}\frac{dp'}{dr}+\frac{g}{\rho c^2_s} P' + (N^2 - \omega^2)
\xi_r = -\frac{d\phi'}{dr},
\end{equation}
and
\begin{equation}
\label{eq:nao3}
\frac{1}{r^2}\frac{d}{dr} \left ( r^2 \frac{d\phi'}{dr} \right ) -
\frac{l(l+1)}{r^2}\phi'=4\pi G \rho \left ( \frac{P'}{\rho c^2_s} +
\frac{N^2}{g}\xi_r \right ).
\end{equation}
$\xi_r$ and $\xi_h$
are the radial and horizontal components of the Lagrangian
displacement vector, $\vec{\xi} = \vec{r} - \vec{r_0}$, and are defined
by
\begin{equation}
\vec{\xi} = \left ( \xi_r,\xi_h \frac{\partial}{\partial \theta},
\xi_h \frac{\partial}{\sin{\theta} \partial \phi}\right ) Y_{lm}
(\theta,\phi)e^{i \omega t}
\end{equation}
and
\begin{equation}
\xi_h = \frac{1}{\omega^2 r} \left( \frac{P'}{\rho} + \phi' \right ).
\end{equation}
$N$ and $L_{\ell}$ are the Brunt-V\"{a}is\"{a}l\"{a} and Lamb
frequencies, respectively, which are given by
\begin{equation}
\label{eq:brunt}
N^2 = g \left ( \frac{1}{\Gamma_1} \frac{dlnP}{dr} - \frac{dln
  \rho}{dr}\right )
\end{equation}
and
\begin{equation}
\label{eq:lamb}
L^2_{\ell}=\frac{\ell(\ell+1)c_s^2}{r^2},
\end{equation}
where $g=GM_r/r^2$, $G$ is the gravitational constant, and $M_r$ is the
mass interior to $r$.

Equations (\ref{eq:nao1}), (\ref{eq:nao2}), and (\ref{eq:nao3}) represent a
fourth-order boundary value problem.  Near $r \sim 0$, simplifications
and regularity in the variables
$\phi'$, $\xi_r$, and $(P'/\rho + \phi')$ dictate that the inner
boundary conditions are
\begin{equation}
\label{eq:ibc1}
\frac{d\phi'}{dr} - \frac{\ell \phi'}{r}=0
\end{equation}
and
\begin{equation}
\label{eq:ibc2}
\xi_r - \ell \xi_h = 0.
\end{equation}
At the surface of the calculational domain, $r=R$ or $M_r=M$, one outer boundary
condition, from the regularity of $\phi'$, is
\begin{equation}
\label{eq:obc1}
\frac{d\phi'}{dr} + \frac{(l+1)}{r}\phi' = 0.
\end{equation}
A second outer boundary condition
allows for the possibility that the atmosphere is either evanescent to
both p- and g-waves or progressive to outgoing p-waves \markcite{lai00c}({Lai} 2000b), depending upon the mode frequency:
\begin{equation}
\label{eq:obc2}
\left (\frac{\lambda_{-} - (V/\Gamma_1-3)}{b_1} \right ) \frac{\xi_r}{r} -
\frac{\omega^2}{g} \xi_h - \left ( \frac{\alpha_1(\lambda_{-} -
  (V/\Gamma_1-3))}{b_1} - \alpha_2 \right ) \frac{\phi'}{g r} =
0,
\end{equation}
where
\begin{equation}
\lambda_{-} = \onehalf \left ( \frac{V}{\Gamma_1} + \frac{rN^2}{g}-2 -
\gamma^{1/2} \right ),
\end{equation}
\begin{equation}
\gamma = \left ( \frac{rN^2}{g}-\frac{V}{\Gamma_1}+4 \right )^2 +
4\left ( \ell(\ell+1)\frac{GM}{\omega^2 R^3}-\frac{V}{\Gamma_1}\right
) \left ( \frac{\omega^2 R^3}{GM}-\frac{rN^2}{g} \right ),
\end{equation}
\begin{equation}
b_1=\frac{\ell(\ell+1)\omega^2 R^3}{GM}-\frac{V}{\Gamma_1},
\end{equation}
\begin{equation}
b_2=\frac{\omega^2 R^3}{G M}-\frac{rN^2}{g},
\end{equation}
\begin{equation}
\alpha_{1}=\frac{b_1(rN^2/g) - ((1+rN^2/g) + \ell)V/\Gamma_1}{(V/\Gamma_1 - 3 +
  \ell)(1+rN^2/g + \ell) - b_1b_2},
\end{equation}
and
\begin{equation}
\alpha_{2}=\frac{b_2V/\Gamma_1 - (V/\Gamma_1 - 3 +
  \ell)rN^2/g}{(V/\Gamma_1 - 3 + \ell)(1+rN^2/g + \ell) - b_1b_2}.
\end{equation}
If the mode frequency is such that this boundary condition represents
leakage of mode energy by progressive p-waves then the eigensolutions
are complex, with the imaginary part of $\omega$, $\omega_I$, being
related to the rate of leakage.

The solutions to the boundary and eigenvalue problem as defined by
eqs. (\ref{eq:nao1})-(\ref{eq:nao3}) with the boundary
conditions, eqs. (\ref{eq:ibc1})-(\ref{eq:obc2}), are the eigenmodes of the perturbed star.  The
eigenvalue, $\omega^2$, and eigensolution depend on the stellar
structure and spherical harmonic index, $\ell$, but without rotation are degenerate for
different values of $m$.  We have found solutions to this boundary value
problem using a shooting technique, in which we search for all modes within a region of the complex plane of $\omega^2$.  This is
accomplished with a 2-D
Newton-Raphson technique, where convergence to one part in $10^6$, or better, in
the eigenvalue is required.  We compare our
eigensolutions and eigenfrequencies for the standard Solar model
\markcite{christensen96}({Christensen-Dalsgaard} {et~al.} 1996) with the eigenmodes determined by a publicly
accessible Fortran code\footnote{provided by J. Christensen-Dalsgaard
  http://astro.phys.au.dk/$\sim$jcd/adipack.n/} designed to solve the
adiabatic linear
perturbation equations.  The agreement between the two
is better than 0.1\%.

\subsection{Work Integral}
\label{sec:work}
In general, non-adiabatic effects such as nuclear burning and
neutrino emission are included in the perturbation equations.  However, if these terms are small,
the contributions to the imaginary
part of $\omega$, $\omega_I$, which is directly related to the rate of
growth or decay of the mode, can be obtained from the time-averaged work integral, $W_j$ \markcite{unno89}({Unno} {et~al.} 1989).  For a
quasi-adiabatic oscillation,
\begin{equation}
\label{eq:work}
W_j = \frac{\pi}{\omega} \int_0^M \epsilon_j \left ( \epsilon_{jT}  +
  \frac{\epsilon_{j\rho}}{\Gamma_3 - 1} \right ) \left | \frac{\delta T}{T}
  Y_{lm}(\theta,\phi) \right |^2 dM_r,
\end{equation}
where $\epsilon_j$ (erg g$^{-1}$ s$^{-1}$) is the deposition or loss
of energy by either nuclear heating if $j = nuc$ or neutrino losses if
$j = \nu$.  Furthermore, $\epsilon_{jT} = (\partial ln \epsilon_j/ \partial ln T)_{\rho}$,
$\epsilon_{j\rho} = (\partial ln \epsilon_j/ \partial ln \rho)_{T}$,
and $\Gamma_3 - 1
\equiv (\partial ln T / \partial \rho)_S$, where $S$ is the entropy.
The radial distribution of the Lagrangian
perturbation in temperature is given by
\begin{equation}
\label{eq:tpert}
\frac{\delta T}{T} = V \nabla_{ad} \left (
\frac{\omega^2}{g} \xi_h - \frac{\phi'}{g r} - \frac{\xi_r}{r} \right ),
\end{equation}
where $\nabla_{ad} = (\partial ln T/ \partial ln \rho)_s$ and $V = gr
\rho/P$.  

When $\omega$ is complex, as is the case when the mode gains or loses
energy, the time dependence of the modes may be written as
\begin{equation}
\vec{\xi} \propto e^{i \omega_R t}e^{t/\tau},
\end {equation}
where $\omega_R$ is the real part of $\omega$, and $\tau$ is the timescale for growth (+) or decay (-) and is a composite
effect determined by the equation
\begin{equation}
\label{eq:tau}
\frac{1}{\tau} = \frac{1}{\tau_{nuc}} + \frac{1}{\tau_{\nu}} +
\frac{1}{\tau_{leak}}.
\end{equation}
$\tau_{nuc}$, $\tau_{\nu}$, and $\tau_{leak}$ include the effects of
nuclear gains, neutrino losses, and acoustic losses, respectively.
The nuclear and neutrino timescales are related to their respective work integrals by
\begin{equation}
\label{eq:tauj}
\tau_j =  \frac{4 \pi E_W}{\omega_R W_j},
\end{equation}
where $E_W$ is the energy in the mode:
\begin{equation}
\label{eq:ew}
E_W = \frac{\omega_R^2}{2}\int^M_0 \left | \vec{\xi} \right |^2 d M_r,
\end{equation}
and $\tau_{leak}$ is directly determined by the imaginary part of
$\omega$, $Im(\omega) = -1/\tau_{leak}$. This quantity is the solution to the adiabatic
linear perturbation equations when eq. (\ref{eq:obc2}) is the fourth boundary condition.
The modes are stable or unstable for $\tau < 0$
or $\tau > 0$, respectively.

\section{Massive Star Structures}
\begin{figure}
\plotone{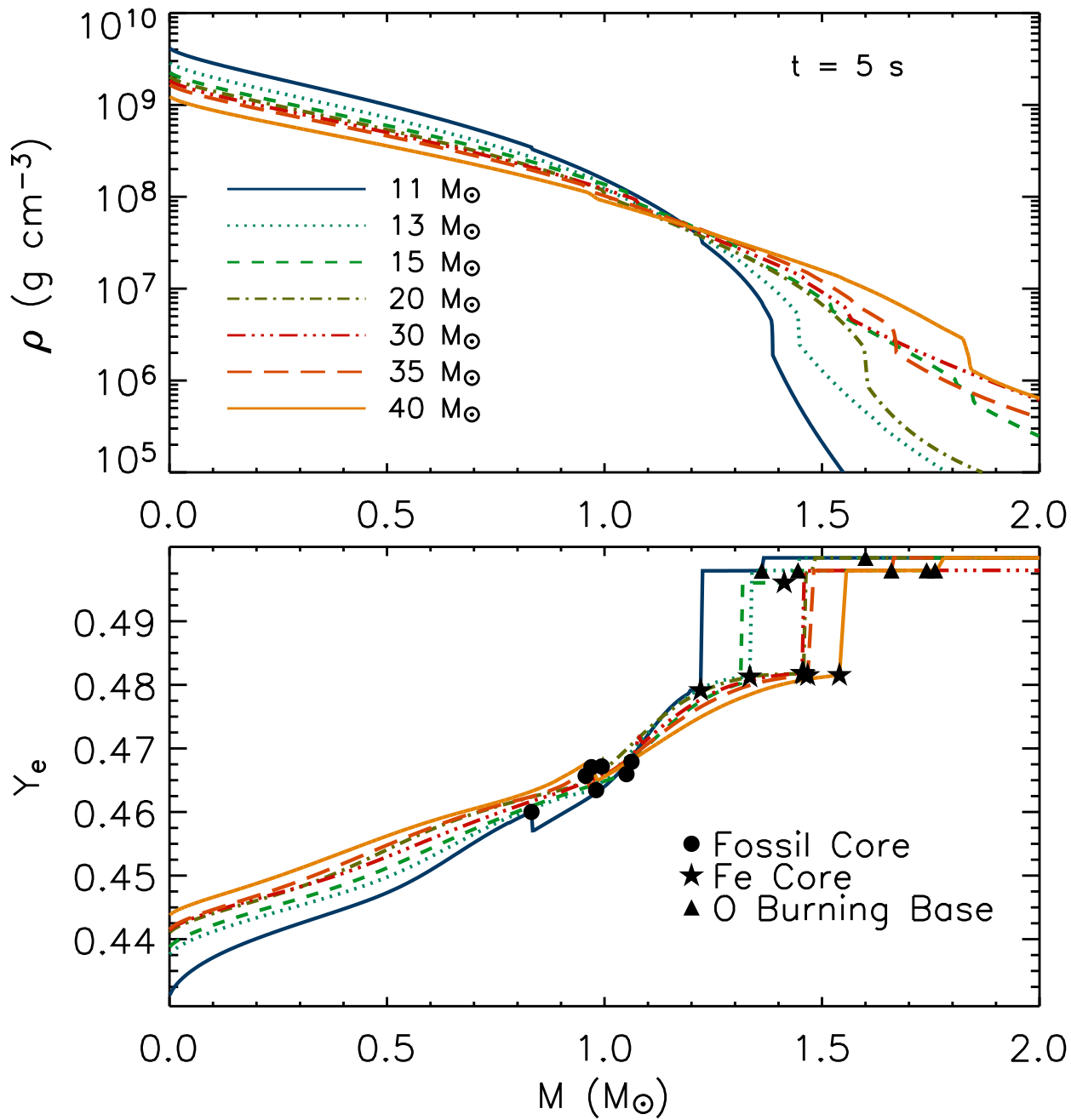}
\caption{Density (top panel) and electron fraction,
  $Y_e$, (bottom panel) versuses interior mass, $M$, for the 11 M$_{\sun}$ (solid and dark
  blue), 13M$_{\sun}$ (dotted and light blue), 15 M$_{\sun}$ 
  (dashed and green), 20 M$_{\sun}$ (dot-dash and olive green), 30
  M$_{\sun}$ (dot-dot-dot-dash and red), 35 M$_{\sun}$ (long dashed
  and orange), and 40 M$_{\sun}$ (solid orange) models, all at 5 s
  before the onset of collapse.  Note in the top panel that the lower mass models have
  higher central densities, but that the higher mass models have
  shallower density gradients.  Included on the $Y_e$ plot are points
  indicating significant features in the star, which greatly affect
  the modes.  Circles represent the farthest extent of the fossil Fe
  core, which was formed during convective core Si burning.  Stars
  represent the base of the Si burning shell at 5 seconds before
  the start of collapse (i.e., the extent of the Fe core, including the fossil Fe and
  the most recent ashes from Si shell burning), which generally aren't
  convective until collapse.  The triangles represent the base of the
  convective O burning shell\label{fig:rhoandye}} 
\end{figure}

To investigate the character of the eigenmodes as a function of
progenitor mass and time, $t$, prior to the onset of collapse, we consider the cores of
nonrotating, massive stars with masses of 11 M$_{\sun}$, 13 M$_{\sun}$,
15 M$_{\sun}$, 25 M$_{\sun}$, 30 M$_{\sun}$, 35 M$_{\sun}$ and 40
M$_{\sun}$ \markcite{rauscher02,woosley02}({Rauscher} {et~al.} 2002; {Woosley}, {Heger}, \& {Weaver} 2002) and at $t = $ 3600, 600, 60, 30,
25, 20, 15, 10, 5, 2, 1, and 0 (start of collapse) seconds before the
onset of collapse.  The commencement of collapse is defined at the time when
the peak infall velocity exceeds 900 km s$^{-1}$.  This is about 200
to 250 ms before core bounce.
The structure of each stellar model uniquely determines its
eigenmodes.  Thus, an
indentification of important trends in the models is helpful in
characterizing these modes.  In the top panel of
Fig. \ref{fig:rhoandye} we see that more massive models have higher entropy,
translating into lower central densities and more extended
density ($\rho$) versus interior mass, $M_r$, profiles.  Consequently,
prominent boundaries which are associated with discontinuities in entropy and
$Y_e$ and which affect the character of the modes are also located at
larger radii for higher-mass progenitors. These discontinuities are the boundary between the fossil Fe core (the residue of
core Si burning) and the most recent Fe ashes from shell Si burning
(circles on Fig. \ref{fig:rhoandye}),
the edge of the Fe core (includes the fossil Fe core and recent Fe
ashes) and the Si burning shell (stars on Fig. \ref{fig:rhoandye}), and
the interface between the Si burning and O burning shells (triangles
on Fig. \ref{fig:rhoandye}).
The general structural profiles and these distinct boundaries (Table
\ref{tab:boundaries}) together determine the eigenfunctions and eigenfrequencies.

\begin{figure}
\plotone{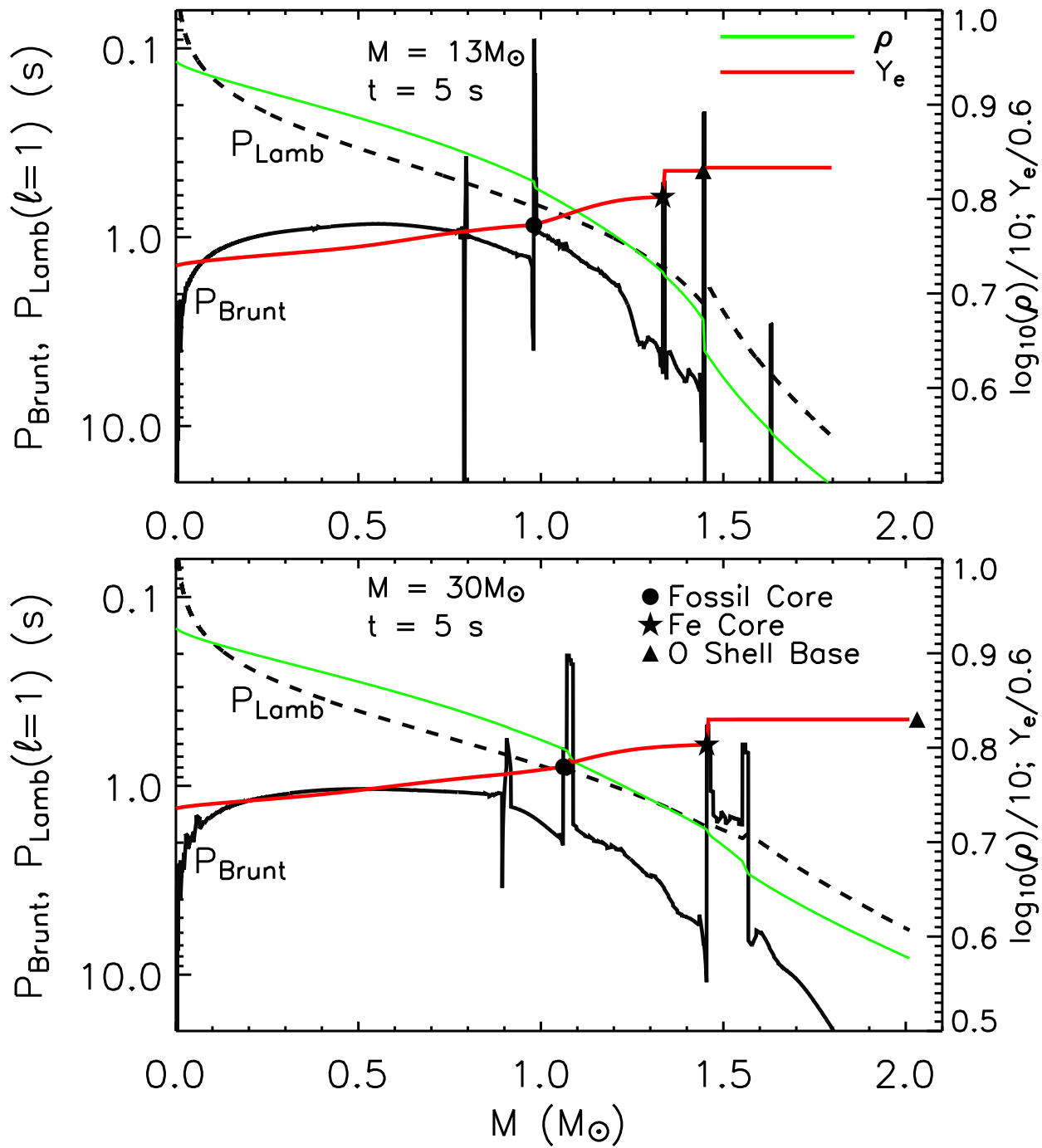}
\caption{Density (solid green) and
  $Y_e$ (solid red) versuses interior mass in comparison to the propagation diagram
  for the 13 M$_{\sun}$ model at 5 s before the onset of collapse.  Usually, the
  square of the Brunt-V\"{a}is\"{a}l\"{a} ($N^2$) and Lamb ($L^2_{\ell}$)
  frequencies are plotted.  In this case, an equivalent period of
  oscillation is displayed instead, defined by
  $P_{Brunt}=2\pi/\sqrt(|N^2|)$ and
  $P_{Lamb}=2\pi/\sqrt(|L^2_{\ell}|)$.  The circles, stars, and triangles
  represent the same structural boundaries as in Fig.
  \ref{fig:rhoandye}.  The vertical spikes are from discontinuous
  density and $Y_e$ changes in the profiles and are real.  The bottom
  panel provides the equivalent information for the 30 M$_{\sun}$
  model.\label{fig:n2andstruct}} 
\end{figure}

The role of structure in affecting the general character of the solutions can be understood using the local
dispersion relation for plane waves of the form $e^{i(\vec{k} \cdot
  \vec{r} \pm \omega t)}$:
\begin{equation}
\label{eq:dispersion}
k^2_r = \frac{(N^2-\omega^2)(L^2_{\ell}-\omega^2)}{c_s^2 \omega^2},
\end{equation}
where $k_r$ is the radial component of the wave vector \markcite{unno89}({Unno} {et~al.} 1989).  This relation
implies that for $N^2 < \omega^2 < L^2_{\ell}$, all waves are
evanescent, for $N^2 < L^2_{\ell} < \omega^2$, they are p-waves, and
for $\omega^2 < N^2 < L^2_{\ell}$, they are g-waves \markcite{unno89}({Unno} {et~al.} 1989).  A propagation
diagram, which plots $N^2$ and $L^2_{\ell}$ versus $M_r$ or $r$,
shows where in the star each type of wave is allowed to
propagate.  The propagation diagrams for the 13
M$_{\sun}$ (top
panel) and 30 M$_{\sun}$ (bottom panel) models and their corresponding
density and $Y_e$ profiles,
are presented in Fig. \ref{fig:n2andstruct}.  The positions of the
prominent discontinuities are marked as in Fig. \ref{fig:rhoandye}.  Rather than plotting $N^2$ and $L^2_{\ell}$, the equivalent
period has been presented for each.  These periods are $P_{Brunt} = 2 \pi /
\sqrt{|N^2|}$ and $P_{Lamb}= 2 \pi / \sqrt{L^2_{\ell}}$.  An
important feature illustrated in Fig. \ref{fig:n2andstruct} is that $N^2$ rises and then falls, creating a
resonant cavity for g-modes.  The fall is partly due to the
density and pressure structures, but on the outside it dramatically plummets
due to convection in the O-burning shell, thereby confining g-waves within the
core.  For higher frequencies (lower periods), p-waves are allowed.
However, they are free to propagate throughout the entire star. Since
the sound-crossing time in the outer stellar mantle is extremely long and dissipative processes
exist, the establishment of standing p-modes is prevented.  Hence, for all models
and times, g-modes are trapped in the core, encompassing the Fe
core and frequently the Si burning shell.

Other salient features evident in Fig. \ref{fig:n2andstruct}
are the spikes in $P_{Brunt}$, or equivalently, $N^2$.  These spikes are
the result of $S$, $Y_e$, and $\rho$ discontinuities at the boundaries
indicated in Table \ref{tab:boundaries}.  As impedance mismatches
these spikes effectively reflect g-waves.  Therefore, even though most g-modes
have their greatest amplitude near the center of the core, the
reflecting boundaries imply that there might be sub-regions within the overall resonating
cavity in which modes may be trapped.

\begin{figure}
\plotone{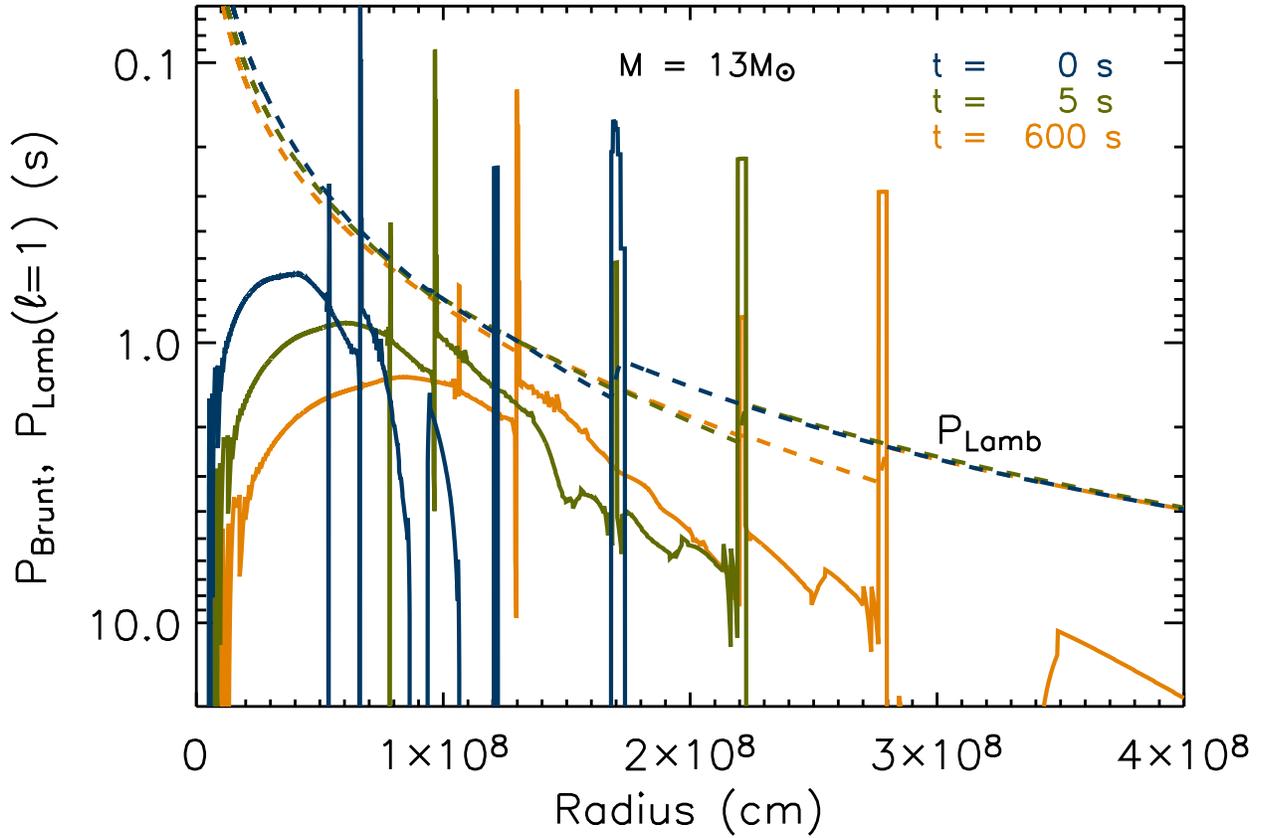}
\caption{A time series of the propagation digram
  for the 13 M$_{\sun}$ model from 600 s to 0 s before the start of collapse.
  $P_{Brunt}$ and $P_{Lamb}$ are defined as in Fig.
  \ref{fig:n2andstruct}.  Notice that the evolution aburptly
  changes for $t = 0$ s.  This is due to the dramatic increase in
  nuclear rates during implosive burning, which causes new regions to
  become convective in the mixing-length calculation.\label{fig:timeprop}}
\end{figure}

By prodigious neutrino emission, the cores of massive stars continue to change
their structures prior to the onset of core collapse \markcite{woosley02}({Woosley} {et~al.} 2002).  Figure \ref{fig:timeprop} illustrates the changes of the propagation
diagram for the 13 M$_{\sun}$ model during the last ten minutes of
evolution.  As the inner regions quickly evolve and contract, the Brunt-V\"{a}is\"{a}l\"{a} frequency increases
with time, gradually increasing the modal frequencies.  In contrast, the model corresponding to
$t=0$ s (onset of collapse) exhibits a dramatic truncation in the size of the
resonating cavity, in that $N^2$ drops to negative values as the
convective region grows due to implosive burning.  However, collapse
happens on a dynamical timescale, and will take 200 to 250 ms.
Given that the convective turnover times and the periods of oscillation
are longer than this, the assumption of steady-state mixing-length convection
is probably violated.  Although for completeness, results for the $t=0$ s
models are presented here, the quantitative results at $t=0$ s should
be viewed with healthy skepticism.  At other times, the structural changes are gradual enough for
our peturbation theory to be reliable.

In this paper, the major sources and sinks of energy considered are
nuclear burning, neutrino losses, and losses by
coupling to outward propagating p-waves \markcite{lai00c}({Lai} 2000b).  Since
losses by radiation is overwhelmed by neutrino emission, perturbations
in the radiative flux is ignored.  Throughout the Fe core, neutrino
losses dominate over energy deposition.  Only in the Si- and O-burning
shells do nuclear reactions allow the $\epsilon$-mechanism to operate.  For
the typical g-mode with its largest amplitudes in the center,
neutrino losses will dominate in eq. (\ref{eq:tau}), resulting in
stability.  However, potentially unstable modes are those that are trapped mostly in the outer
part of the core near or in the Si-burning region by the
discontinuities in entropy, $Y_e$, and density \markcite{lai00c}({Lai} 2000b).

\section{Results}
\begin{figure}
\plotone{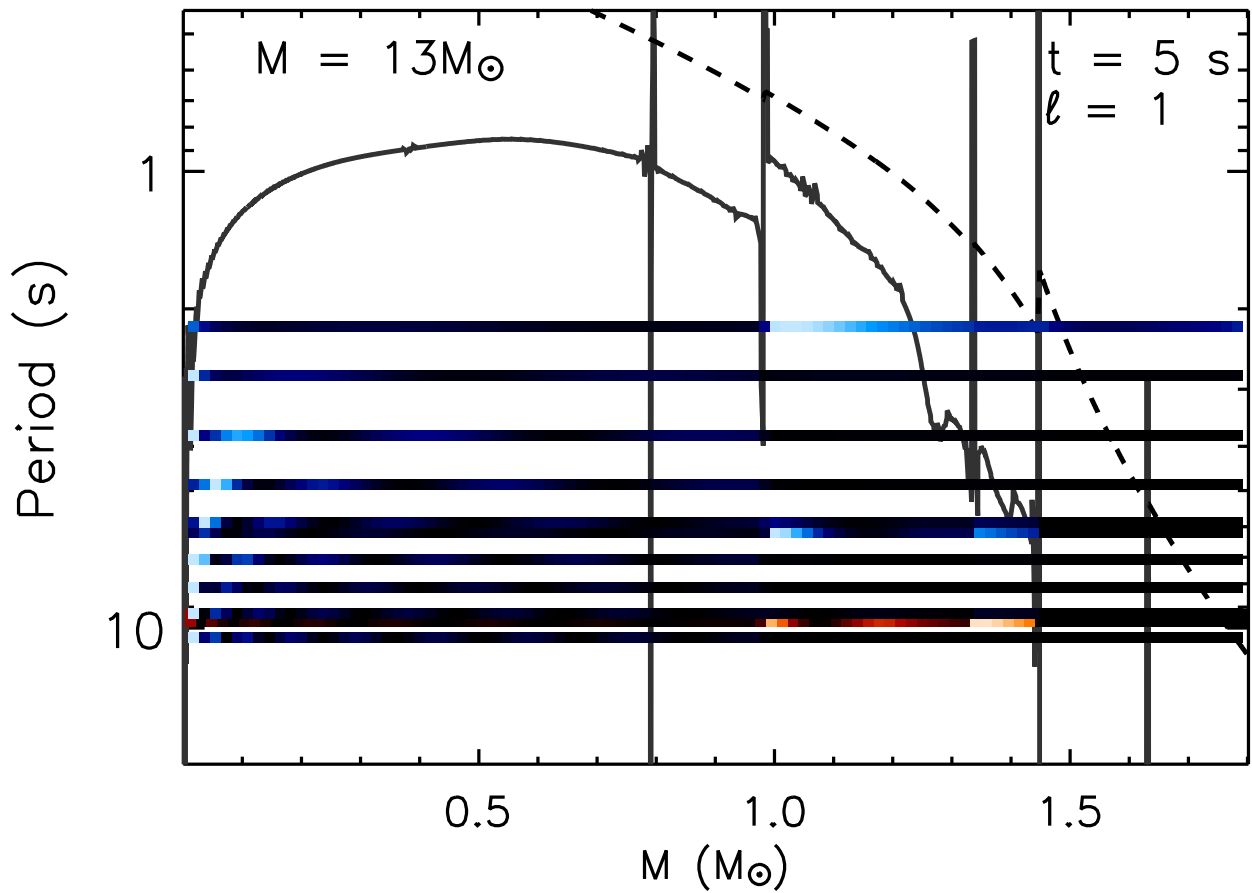}
\caption{The propagation diagram for $\ell=1$ and the 13
  M$_{\sun}$ model at 5 s before the onset of collapse with $P_{Brunt}$ (solid) and
  $P_{Lamb}(\ell=1)$ (dashed), as defined in Fig. \ref{fig:n2andstruct}.  The
  horizontal lines correspond to each eigenmode and period and are
  quantized.  The shading in the lines is a measure of the magnitudes
  of the square of the amplitude, $|\xi|^2$.  Within a single line
  brighter portions indicate a higher amplitude while the darker color
  signifies a lower amplitude.  The blues indicate modes which are
  stable ($\tau < 0$), and the reds correspond to the modes which are
  unstable ($\tau > 0$).\label{fig:propamp13}} 
\end{figure}

\begin{figure}
\plotone{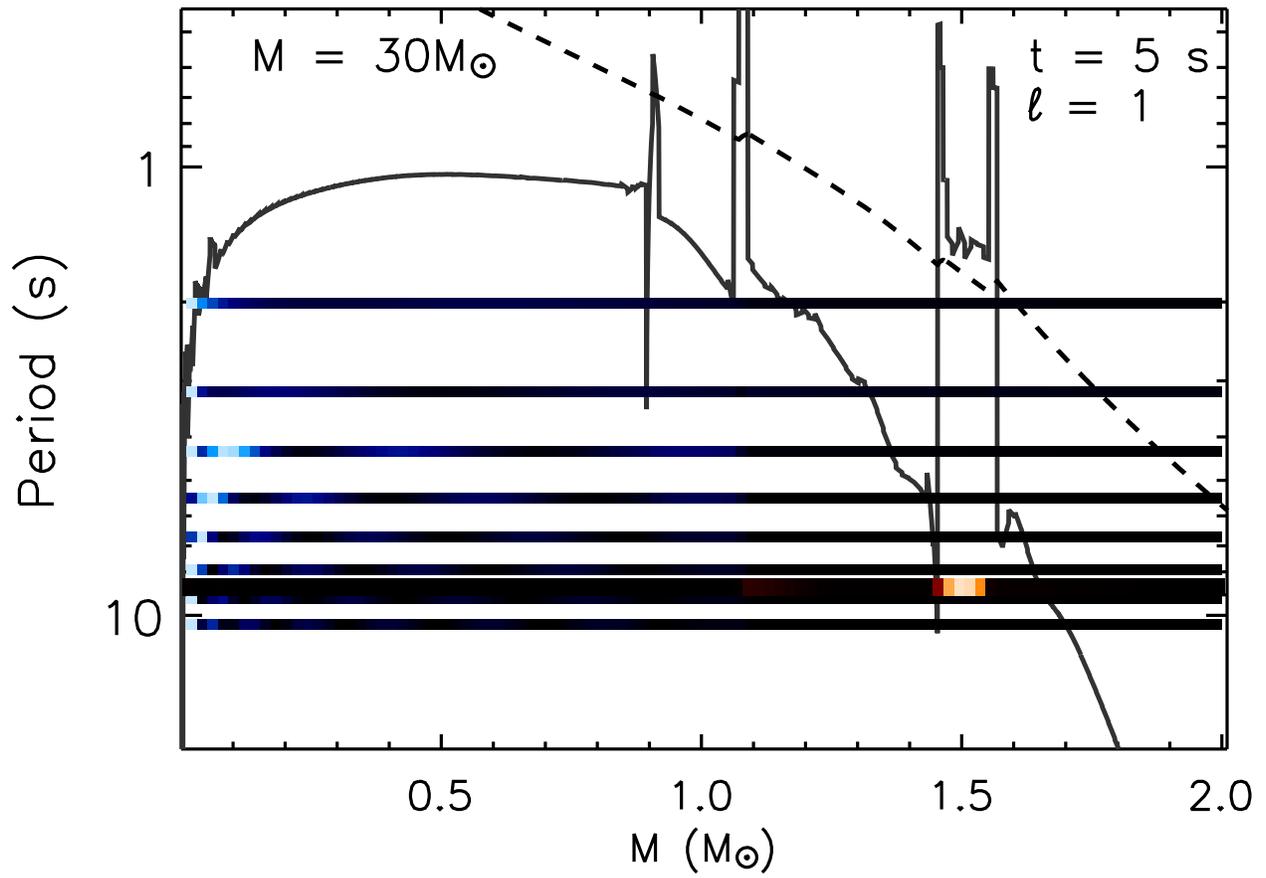}
\caption{This figure is similar to Fig.
  \ref{fig:propamp13}, but for the 30 M$_{\sun}$
  model.\label{fig:propamp30}} 
\end{figure}

\begin{figure}
\plotone{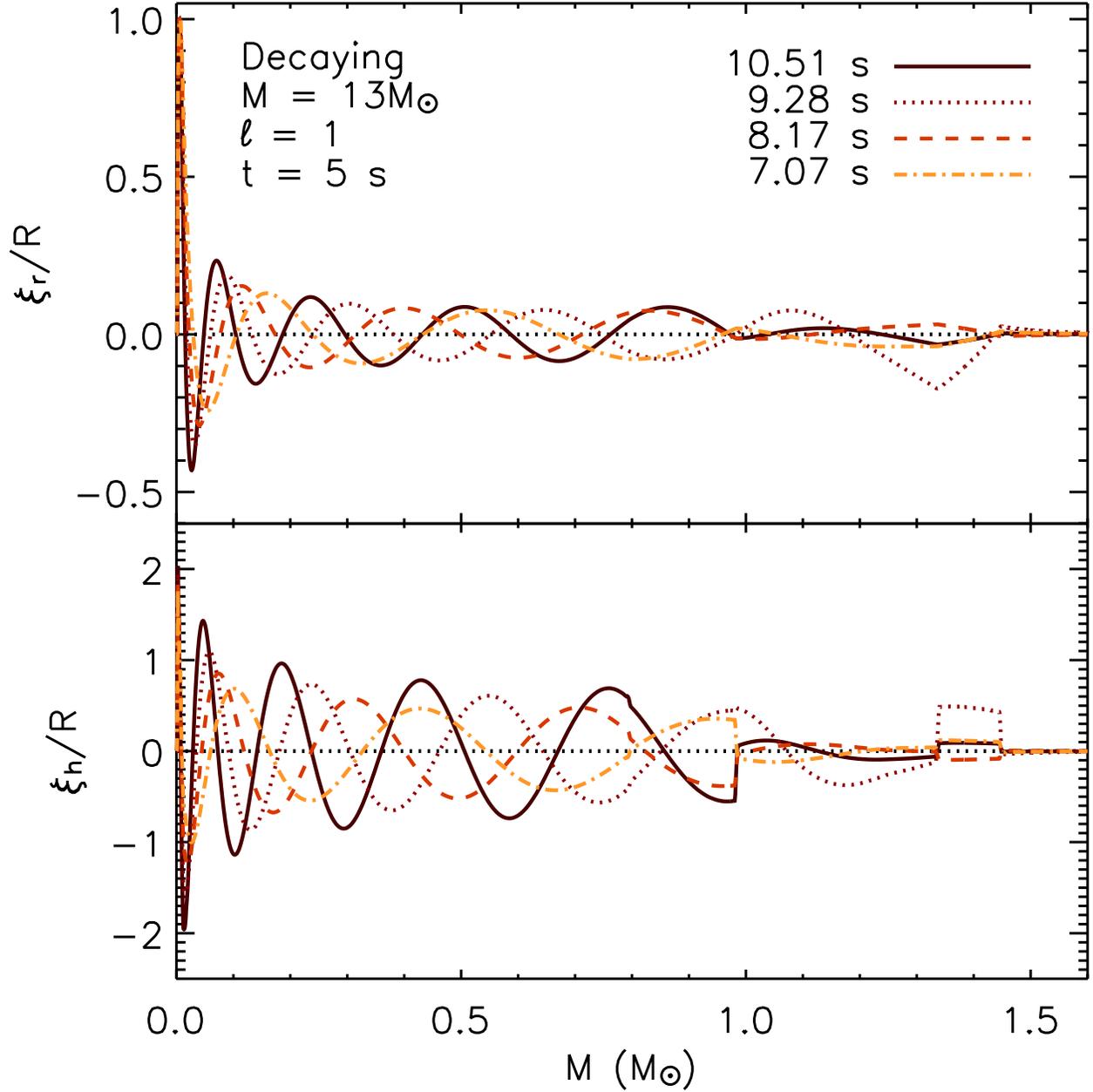}
\caption{The dimensionless eigenfunctions of {\it stable}
  g-mode oscillations with periods of 10.51, 9.28, 8.17, and 7.07 s,
  for $\ell=1$, 13 M$_{\sun}$, and 5 s before the start of
  collapse.  Each mode is distinguished by its period.  The top
  panel plots $\xi_r/R$, where $\xi_r$ is the radial component of
  perturbation and $R$ is the calculation domain size.  The amplitude
  is arbitrarily scaled to 1.0 at its peak value.  The bottom panel
  plots the corresponding horizontal amplitude, $\xi_h/R$.\label{fig:13decay}}
\end{figure}

\begin{figure}
\plotone{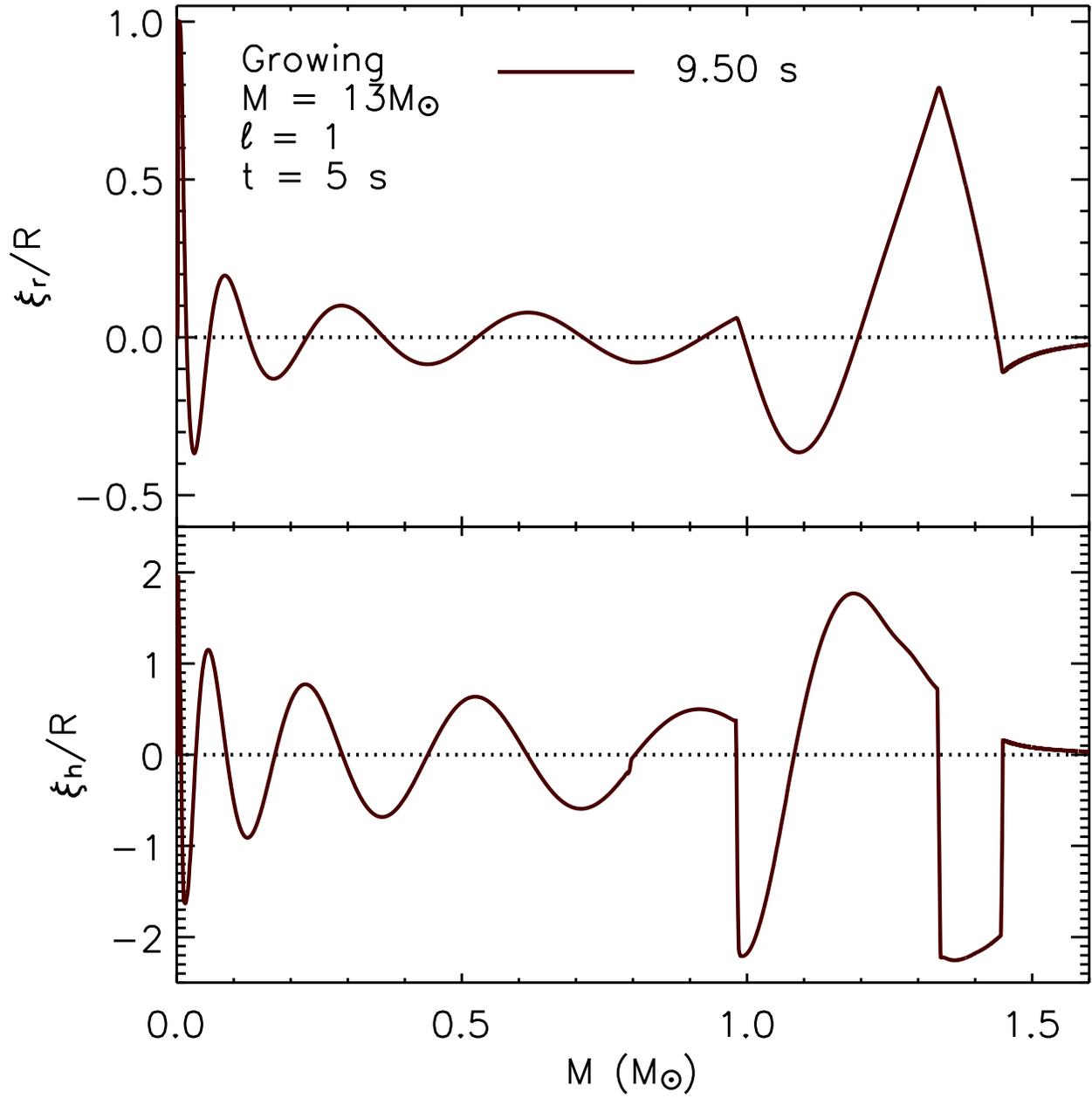}
\caption{Plotted are the same attributes depicted in Fig. \ref{fig:13decay},
  but for a representative {\it unstable} mode with period 9.50 s.\label{fig:13grow}}
\end{figure}

\begin{figure}
\plotone{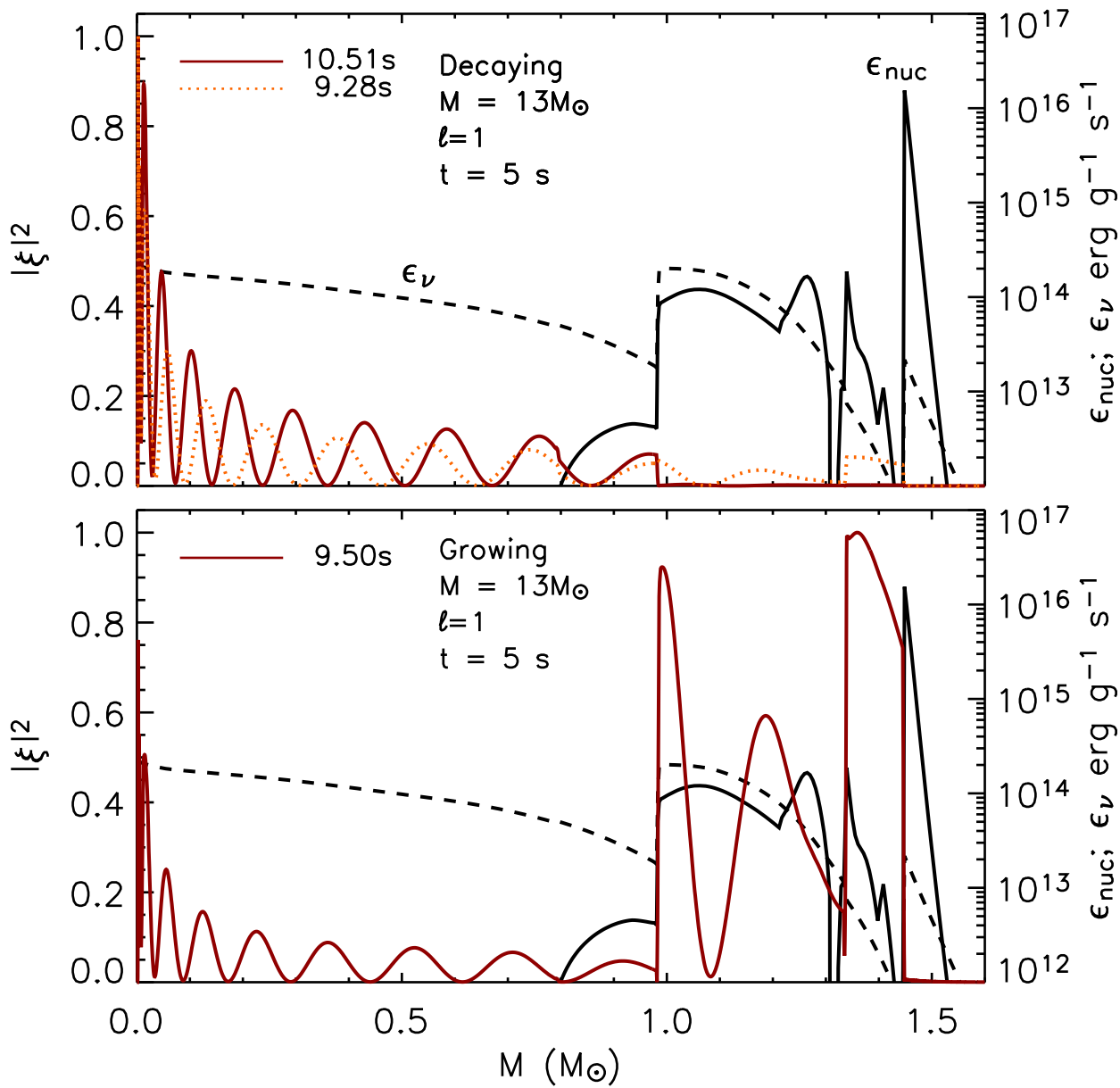}
\caption{The Lagrangian displacement squared, $|\xi|^2$ (browns and oranges), the
  local deposition of energy by nuclear processes, $\epsilon_{nuc}$
  (solid and black), and the
  local losses of energy due to neutrinos, $\epsilon_{\nu}$ (dashed),
  versus interior mass (M$_{\sun}$) for the 13 M$_{\sun}$ model,
  $\ell=1$, and 5 s before the onset of collapse. The top panel illustrates this
  comparison for two representative stable modes (periods = 10.51 and
  9.28 s), while the bottom panel plots an unstable
  mode (period = 9.5 s).  Note that the inner-core g-modes have their
  largest relative amplitudes where neutrino losses dominate, but that
  the outer-core g-modes have their largest relative amplitudes
  trapped in the recent Si burning ashes as well as the Si burning
  layer giving rise to instability.\label{fig:rates13}} 
\end{figure}

\begin{figure}
\plotone{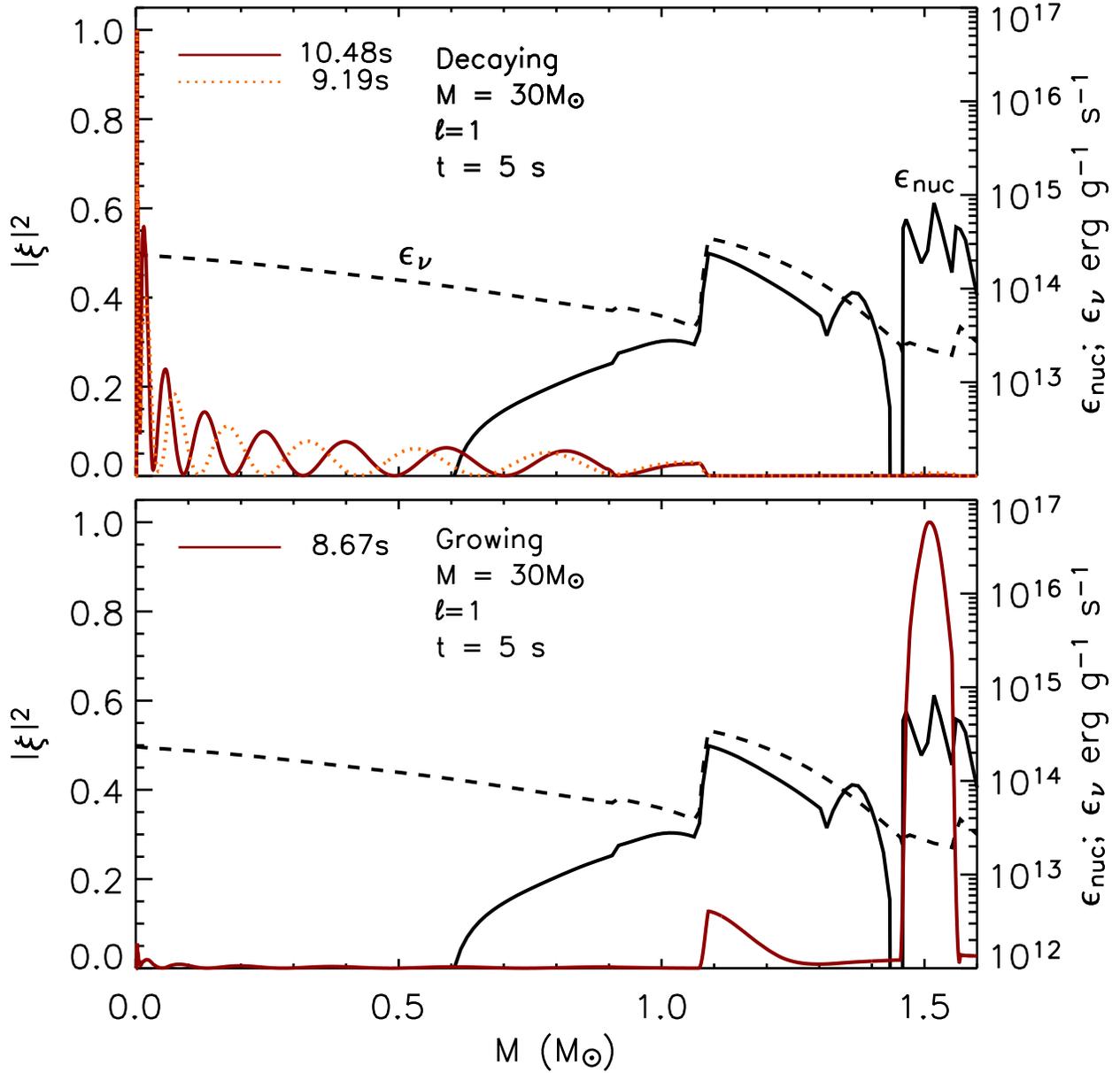}
\caption{Same as Fig. \ref{fig:rates13}, but for the $M = 30$
  M$_{\sun}$ progenitor, stable modes with periods of 10.48 s and
  9.19 s, and an unstable mode with period = 8.67 s.\label{fig:rates30}} 
\end{figure}

\begin{figure}
\plotone{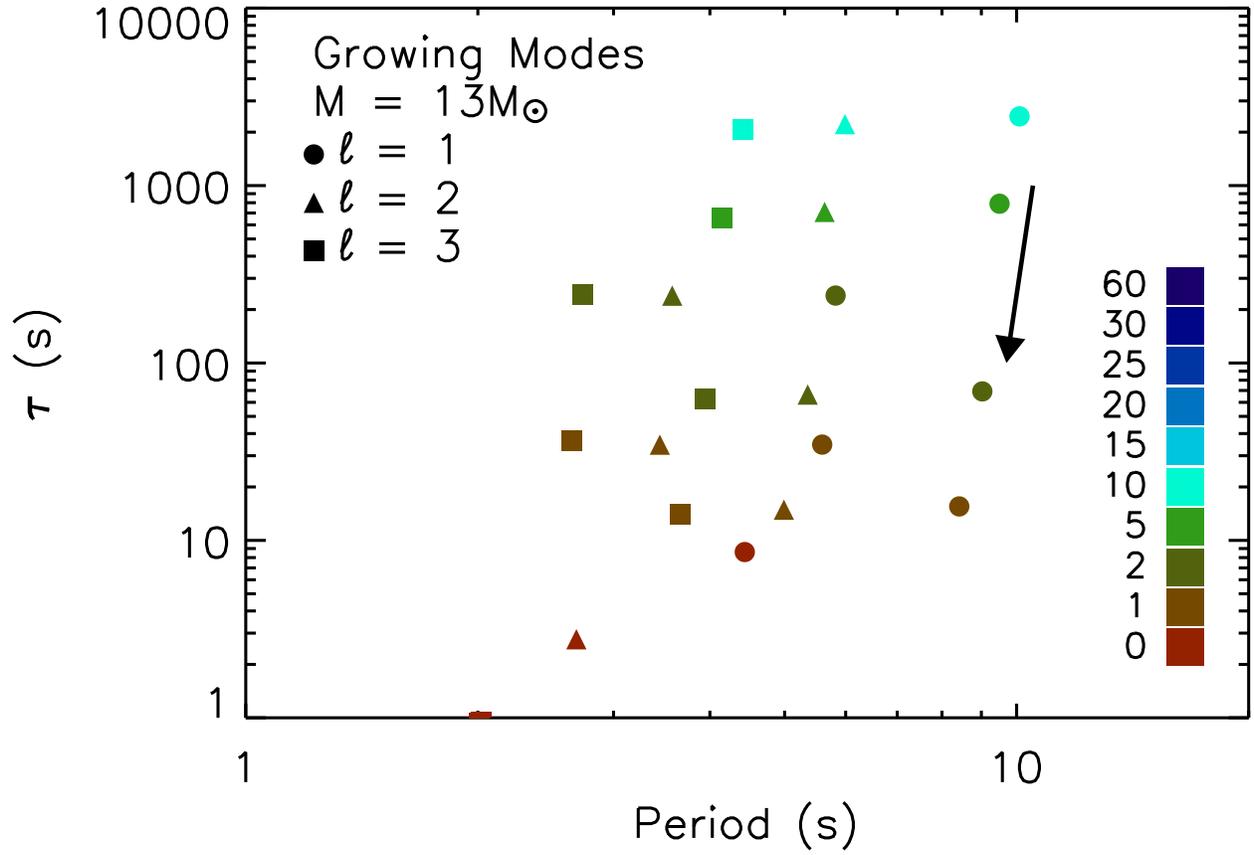}
\caption{The growth timescales, $\tau$, in seconds, for
  the unstable modes versus their periods, in seconds, for the 13
  M$_{\sun}$ model.  The plots are for $\ell =$ 1 (circle), 2 (triangle), and 3
  (square) and for times of 60 (blue), 30,
  25, 20, 15, 10, 5, 2, 1, and 0 (red) seconds before the onset of collapse.  The arrow
  indicates the evolution of the period and growth timescales for
  particular modes as the models approach collapse.\label{fig:tau13}} 
\end{figure}

\begin{figure}
\plotone{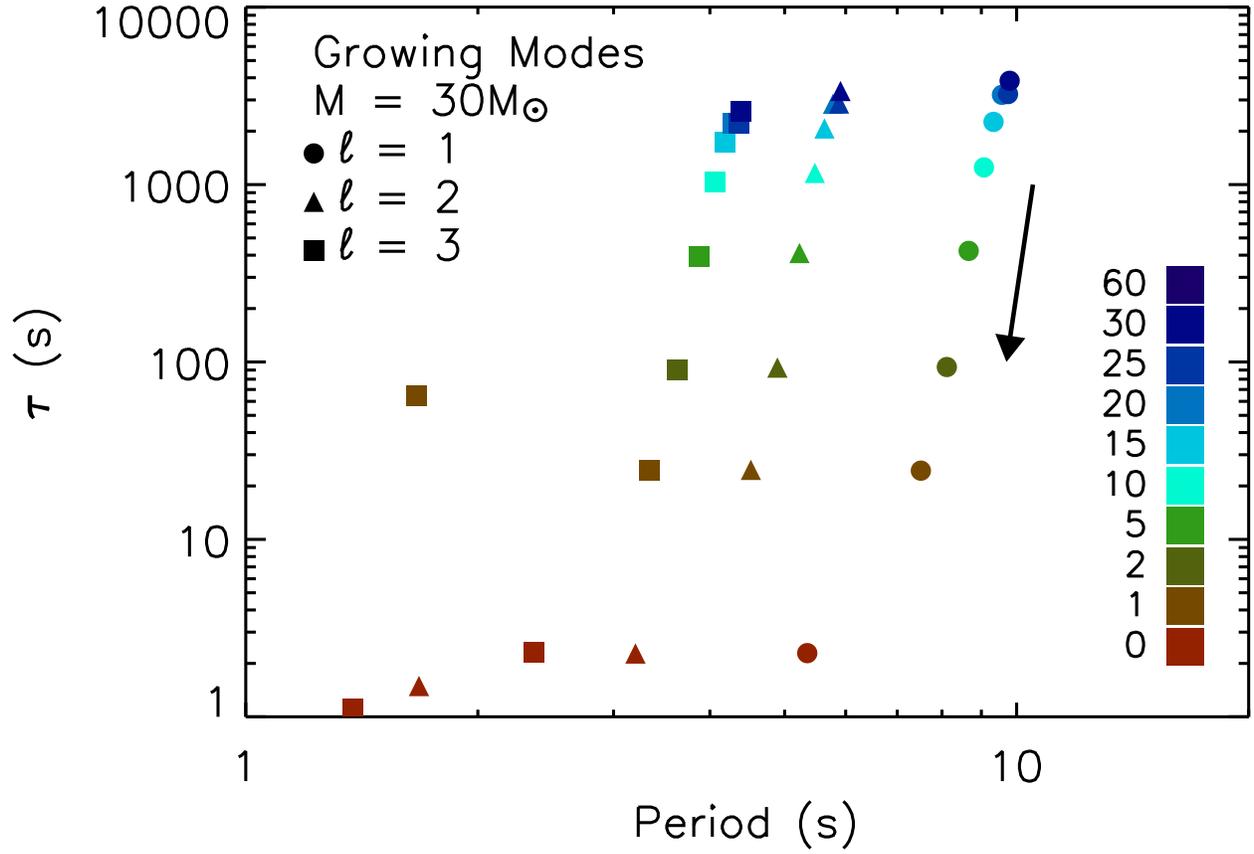}
\caption{Same as Fig. \ref{fig:tau13}, but for the $M = 30$
  M$_{\sun}$ progenitor.  Note  the clear, uninterupted evolution of
  specific g-modes up to the onset of collapse.\label{fig:tau30}}
\end{figure}

The results of our analysis are summarized in Table
\ref{tab:growing}, which lists all unstable modes found, and
Figs. \ref{fig:propamp13}-\ref{fig:tau30}, which are described in detail below.  Even though we did find modes for
$\ell=$ 1, 2, 3, and higher $\ell$s, we present the results only for $\ell=$ 1, 2, and 3 in Table \ref{tab:growing}.
since the trends at higher $\ell$s are adequatly demonstrated with the
set presented in this paper.  In general, the most interesting
spherical harmonics to
consider are those with odd $\ell$, or those with a top-bottom
asymmetry.  However even $\ell$s are interesting as well, as the nonlinear evolution of the perturbations
into a kick may be triggered by any low order $\ell$.  Since the dipole, $\ell=1$, spherical harmonic
most closely resembles the desired kick geometry, we focus in this
section on $\ell=1$, with peripheral mention of $\ell = 2$ and 3.

Stable and unstable modes are found in all progenitor models.
Figures \ref{fig:propamp13} and \ref{fig:propamp30} display the
propagation diagram for the 13
M$_{\sun}$ and 30 M$_{\sun}$ models, at $t=5$ s before
the commencement of collapse. These diagrams represent modes of the compact and extended
profiles, respectively.  Note again, the spikes resulting from $Y_e$
and $\rho $ discontinuities in $P_{Brunt}$ and the existence of a resonance cavity
confined by O-shell convection in the 13 M$_{\sun}$ model
(Fig. \ref{fig:propamp13}).
Because the O shell is farther out in mass for the more massive 30
M$_{\sun}$ model (Fig. \ref{fig:propamp30}), the density and
pressure profiles are more relevant in defining the resonant region.  In
both figures each horizontal line corresponds to a specific mode;
the shading encodes $|\vec{\xi}|^2$.  Dark and lighter shading signify
low and high displacement, respectively, while blues denote stable
modes and reds identify the unstable modes.
Figures \ref{fig:propamp13} and \ref{fig:propamp30} show that there are
many more stable modes than unstable modes, which is true for all
models at all times and for all $\ell$s.  We also identify two
classes of g-modes, inner-core and outer-core g-modes.  Inner-core g-modes have their greatest
displacements toward the center and diminish with radius.  Their spacing
in $\log_{10}(Period)$ systematically decreases with increasing period
and increasing number of half-wavelengths, $n$.  In contrast
outer-core g-modes do {\it not} have their greatest amplitude in the
center.  They are trapped by discontinuites between the fossil Fe core
and the O shell in the case of the 13 M$_{\sun}$ model and within the
Si burning shell in the case of the 30 M$_{\sun}$ model.  Similar
trapping exists for all other progenitors we consider.

Less than half of the outer-core g-modes are modes
trapped entirely under the spikes (i.e. surface g-modes).  While these
modes satisfy the adiabatic linear oscillation equations, their
physical significance is not well constrained.  Their eigenfunctions and
eigenfrequencies depend on the spike widths and heights, which in turn
depend upon the interface between convective and radiative regions
in mixing-length theory.  However, the results and
conclusions we present in this paper are not predicated on their
character or existence.

A sample of inner-core g-modes for the 13 M$_{\sun}$, $t=5$ s, model
with $\ell = 1$ is plotted versus interior mass in
Fig. \ref{fig:13decay}.  The decreasing envelope of the eigenfunction and
systematic increase in period with increasing half-wavelengths, $n$,
is well explained by a WKB analysis \markcite{unno89}({Unno} {et~al.} 1989).  The
quantization condition, $n \pi = \int k_r dr$, integrated over an
appropriate resonant cavity, for g-waves gives
\begin{equation}
\label{eq:wkbomega}
\omega_R = \frac{[\ell(\ell+1)]^{1/2}}{n \pi} \int_{cavity} \frac{N}{r} dr.
\end{equation}
Numerically, the inner core frequencies are proportional to
$[\ell(\ell+1)]^{1/2}/n$, and integrating eq. (\ref{eq:wkbomega}) over
the appropriate resonance cavity gives approximately the same value.
Following the conservation of wave flux arguments in
\markcite{goldreich99}{Goldreich} \& {Wu} (1999), the envelope for the aribtrarily scaled amplitudes of $\xi_r/R$ and
$\xi_h/R$ in Fig. \ref{fig:13decay} also correspond with those from
the WKB
analysis.  In contrast, a simple WKB analysis fails to predict either the
shape or the period of the outer-core g-mode shown in Fig
\ref{fig:13grow}.  Instead, this mode's period and large displacement farther out in mass
are determined by both resonance in the overall cavity and
resonant trapping between disontinuities in density, $Y_e$, and entropy.

In the context of losses and gains in energy,
Fig. \ref{fig:rates13} displays $|\vec{\xi}|^2$ versus interior mass
for two representative inner-core g-modes (top panel) and an
outer-core g-mode (bottom panel) for the 13 $M_{\sun}$, $t=5$ s model
for $\ell =1$, in addition to $\epsilon_{nuc}$ and $\epsilon_{\nu}$ in erg g$^{-1}$
s$^{-1}$.  Figure \ref{fig:rates30} presents similar information for
the 30 M$_{\sun}$ model.  The overlap of $|\vec{\xi}|^2$ with
$\epsilon_{nuc}$ and $\epsilon_{\nu}$ is an excellent diagnostic proxy for the
integrand in the work integral (eq. (\ref{eq:work})).  Clearly, neutrino
losses dominate the work integral for the case of inner-core g-modes
in both models.  For the particular outer-core g-modes plotted for the
13 M$_{\sun}$  and 30 $M_{\sun}$ progenitors in the bottom panels of
Figs. \ref{fig:rates13} and \ref{fig:rates30}, respectively, the large
amplitude in the Si-burning region allows for instability.
While all outer-core g-modes in
Fig. \ref{fig:propamp30} are unstable, two of the three 13 M$_{\sun}$
outer-core g-modes in Fig. \ref{fig:propamp13} are stablized by neutrino
and acoustic losses.  In general, at least one outer-core g-mode has
the requisite radial perturbation distribution to be unstable.

\begin{figure}
\plotone{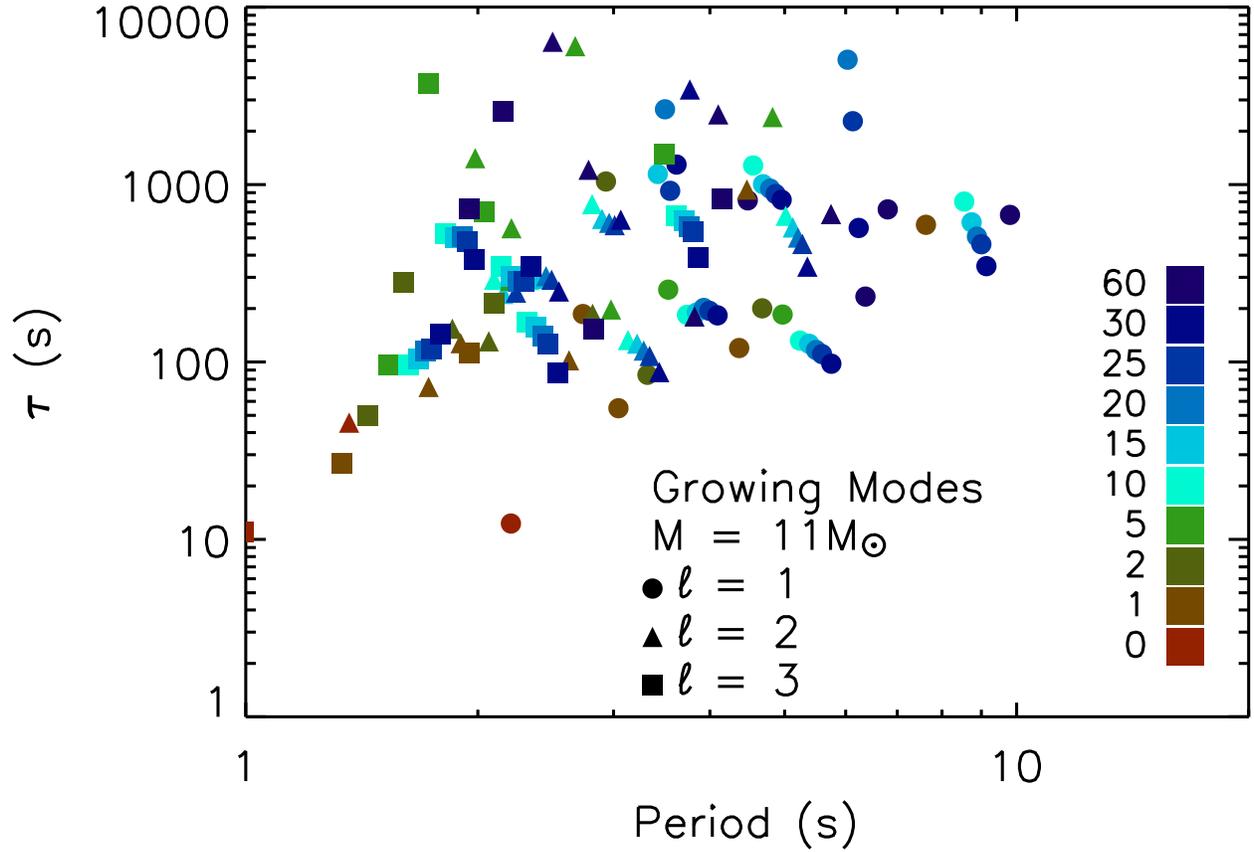}
\caption{Same as Fig. \ref{fig:tau13} but for the $M = 11$
  M$_{\sun}$ progenitor.  While plenty of growing modes exist, their
  timescales are far too long for any significant growth to
  occur.\label{fig:tau11}}
\end{figure}

The growth timescales versus period for $\ell=$ 1, 2, and 3, $t=$ 60,
30, 25, 20, 15, 10, 5, 2, 1, and 0 s before the onset of collapse are plotted for
11 M$_{\sun}$ 13 M$_{\sun}$, and 30 M$_{\sun}$ models in
Figs. \ref{fig:tau11}, \ref{fig:tau13}, and \ref{fig:tau30}, respectively.
As expected, these figures illustrate that for higher $\ell$ the periods
are smaller.  Even though we find many unstable modes for the 11
M$_{\sun}$ we do not find clear trends in the $\tau$ versus period
plane.  In contrast, as represented by the 13 M$_{\sun}$ and 30
M$_{\sun}$ models, all other masses we consider demonstrate a
distinct trend.  The arrows in Figs. \ref{fig:tau13} and \ref{fig:tau30}
indicate the direction in which the modes evolve with time in the
$\tau$ versus period plane.  Specifically, later models are more
compact and have shorter
period oscillations.  In addition, the later, more compact, models have
slightly higher densities and temperatures in the burning regions, thereby
dramatically increasing the burning rates and decreasing $\tau$ in
Figs. \ref{fig:tau13} and \ref{fig:tau30}.  Also note that at no
point in any of the models are the growth timescales shorter than the times until the start of
collapse.
These trends continue to the onset of collapse, which is marked on Figs. \ref{fig:tau13} and
\ref{fig:tau30} with red symbols.  In fact, for the latest models,
$t=0$ s and 1 s before the start of collapse, $\tau$ becomes
comparable to the period, calling into question
the assumption in \S\ref{sec:work} of small nonadiabatic effects.
While dramatic changes imposed by dynamical collapse and implosive
burning for the $t=0$ s models make the effects listed above more
pronounced, the qualitative trends highlighted still obtain for the $t=0$ s models.

For all unstable modes found Table \ref{tab:growing} lists the
progenitor mass, time until onset of collapse, the period, growth
timescale, $\tau$, and the contributions, $\tau_{nuc}$, $\tau_{\nu}$,
and $\tau_{leak}$.  From tens of minutes before
commencement of collapse until collapse itself, unstable modes exist in {\it all} progenitor
models.  However, at every evolutionary stage the timescale for growth
is an order-of-magnitude or two longer than the time until onset
of collapse.
The rates of growth or decay we derive are too long compared to the
time to onset of collapse to
significantly alter the mode amplitudes.  It seems that even though
the $\epsilon$-mechanism is present in the later evolutionary stages
of the cores of massive stars, the timescales imply that its effect
is not large. 

\subsection{Analytic Estimates}
\label{sec:analytic}

Consistently, we find that the timescales for growth are longer than
the time until the onset of collapse.  As such the
$\epsilon$-mechanism is not able to significantly affect the amplitudes
of the unstable g-modes.  Hence, our evaluation of the succesfulness
of the $\epsilon$-mechanism depends upon the validity of the timescale
calculations.  In this section we corroborate,
with an analytic estimate for $\tau_{nuc}$, our numerical
evaluations for the growth timescales.

$W_j$, given by eq. (\ref{eq:work}), is the rate of energy gained or
lost, and $E_W$, eq. (\ref{eq:ew}), is the energy in the mode.
Assuming the region where nuclear burning interacts with a growing
mode is small and the mode itself may be represented by
a top-hat function located entirely and only within
the burning region with maximum displacement $\Delta r/r$, then the integrand
in eq. (\ref{eq:work}) is roughly constant. Therefore,
\begin{equation}
\label{eq:workapprox}
W_j \sim \frac{\pi}{\omega_R} \epsilon_j \epsilon_{jT}  \left ( \frac{\delta
  T}{T} \right )^2 \frac{\Delta M}{4 \pi}, 
\end{equation}
and eq. (\ref{eq:tpert}), simplifies to
\begin{equation}
\frac{\delta T}{T} \sim V \nabla_{ad} \left (
\frac{\Delta r}{r}\right ).
\end{equation}
As a harmonic oscillation, g-modes are the oscillations of
isodensity contours about their equilibrium positions in hydrostatic
equilibrium.  The harmonic potential is $U \sim \onehalf \partial^2
\phi_{eff} / \partial r^2 x^2$, where $x = \Delta r e^{i \omega t}$.  From $\vec{\nabla} \phi = \vec{\nabla} P
/ \rho$, we get $\phi_{eff} \sim \phi \sim P/\rho \sim
GM_r/r$. Hence, the potential energy, $u$, is given by
\begin{equation}
u \sim \frac{GM_r}{r} \left ( \frac{x}{r}\right )^2.
\end{equation}
Since $E_W = 2 \int <u> dM_r$, where $<>$ denotes the time average
over one period of oscillation, then
\begin{equation}
\label{eq:ewapprox}
E_W \sim \frac{GM_r}{r} \left ( \frac{\Delta r}{r} \right )^2
\frac{\Delta M}{4 \pi}.
\end{equation}
Finally, taking the ratio of eq. (\ref{eq:ewapprox}) and eq. (\ref{eq:workapprox})
gives a rough estimate for $\tau$:
\begin{equation}
\label{eq:tauapprox}
\tau_{nuc} = \frac{4 \pi E_W}{\omega_R W_{nuc}}\sim 4 \frac{P/(\rho V)}{\epsilon_{nuc} \epsilon_{nucT} \nabla_{ad}^2},
\end{equation}
In the burning region, for the 30 M$_{\sun}$ $t=5$ s model and the 8.67 s, $\ell =1$ mode, $P/\rho
\sim 2.2 \times 10^{17}$ erg g$^{-1}$, $\epsilon_{nuc} \epsilon_{nucT} \sim 2
\times 10 ^{16}$ erg g$^{-1}$ s$^{-1}$, $\nabla_{ad} \sim 1/4$, and $V
\sim 2$, giving $\tau_{nuc} \sim 320$ s from eq. (\ref{eq:tauapprox}).  The numerical
solution in Table \ref{tab:growing} is $\tau_{nuc}
= 328$ s.  Similarly for the 8.45 s, $\ell=1$ mode of the 13 M$_{\sun}$
progenitor model at $t=1$ s, $P/\rho \sim 1.4 \times 10^{17}$ erg g$^{-1}$,
$\epsilon_{nuc} \epsilon_{nucT} \sim 10 ^{17}$ erg g$^{-1}$ s$^{-1}$,
$\nabla_{ad} \sim 1/4$, and $V \sim 3$, giving 
$\tau_{nuc} \sim 30$ s from eq.(\ref{eq:tauapprox}).
The numerical solution (Table \ref{tab:growing}) is $\tau_{nuc} = 15.5$ s.  That the simple
analytic arguments and estimates approximately reproduce the numerical
results is reassuring.  Even though the
nuclear rates increase toward later times, at no time is this trend enough to
cause significant growth prior to collapse.

\section{Summary and Conclusions}
For all progenitors, we find stable and unstable g-modes with
oscillation periods in the range from $\sim$1 to $\sim$10 seconds.  The stable
g-modes are often concentrated in the inner core and are stabilized by neutrino
emission.  Unstable g-modes on the other hand are trapped in the
outer radii of the core by discontinuities in $S$, density, and $Y_e$.
Their typical growth timescales determined numerically and
analytically range from 10s to 10,000s of seconds, which
are long compared to the time until the start of collapse.  Therefore,
we conclude that the $\epsilon$-mechanism is an unlikely
source for large perturbations in the progenitor prior to the onset of
collapse.

There are several caveats to our conclusion.  Our results are for a set of 1-D nonrotating models from
\markcite{rauscher02}{Rauscher} {et~al.} (2002) and \markcite{woosley02}{Woosley} {et~al.} (2002).  Rotation in the late stages of massive stellar
evolution may
significantly effect the structure of the star prior to collapse
\markcite{heger00,heger03}({Heger}, {Langer}, \& {Woosley} 2000; {Heger}, {Woosley}, \& {Langer} 2003), thereby altering the
analysis in this paper.  However, with modest Fe core rotation periods
($\ga 10$ s) we expect little deviation from our results, and the
simple energy arguments in \S\ref{sec:analytic} should hold.  We also have not considered a
different suite of progenitor models \markcite{nomoto88,limongi03}({Nomoto} \& {Hashimoto} 1988; {Limongi} \& {Chieffi} 2003), which may have different
eigenmodes.  In addition, calculating full 3-D, non-linear, dynamical
convection with nuclear reactions during
the final stages of massive star evolution is a task for the future;
the progenitors and their g-modes may be significantly different than
assumed or calculated here.

We also have not considered coupling
between convective modes and g-modes, which provides an excellent
mechanism for the excitation of the 5-minute p-modes in the Sun \markcite{goldreich90}({Goldreich} \& {Kumar} 1990).
In such analyses, the amplitudes of the modes are estimated by
assuming steady state is achieved in the nonlinear time-dependent
amplitude equations.  However, the short timescales until the onset of
collapse would violate the assumption of steady state and invalidate
such analyses.  More importantly, it would be unlikely that the short
timescales until the onset of collapse would allow for significant
excitation of g-modes by convection.

On the other hand, the growth of perturbations in the supersonic regions during
collapse does occur \markcite{lai00a,lai00b}({Lai} 2000a; {Lai} \& {Goldreich} 2000). Because we have not ruled out all possible
sources of perturbations prior to collapse, asymmetries in the
progenitor may still be relevant in producing the largest kicks.  For example, convection in
the last stages of evolution is often quite vigorous and may have Mach
numbers in the range 0.1 to 0.2 and density perturbations, $\delta
\rho/\rho$, in the range 0.01 to 0.05 \markcite{bazan98}({Bazan} \& {Arnett} 1998).  These
perturbations may indeed be sufficient to produce the observed kicks \markcite{burrows96}({Burrows} \& {Hayes} 1996).

In performing a stellar pulsational and stability analysis of the
late-time cores of massive stars, we have identified many stable and unstable
modes.  We note that stability is predominantly determined by the
neutrino losses or energy deposition by nuclear reactions.  Compared
with the time until the onset of collapse, the timescales for growth and decay are
long.   In \S\ref{sec:analytic}, we argue that the long growth
timescales (compared with the time until the start of collapse) are
confirmed by energetic arguments.  We suggest that
the $\epsilon$-mechanism for g-mode growth does not generate the requisite perturbations
prior to collapse needed to stimulate large kicks during the supernova explosion.  Therefore, we
conclude that if the hydrodynamic mechanism in combination with
progenitor perturbations is to succeed in explaining neutron star
kicks other sources of seed perturbations must be investigated.

\acknowledgments

We thank Phil Arras, Martin
Pessah, Tony Piro, Casey Meakin, Stan Woosley, Dong Lai, and Chris Fryer for helpful discussions.
Support for this work is provided in part by
the Scientific Discovery through Advanced Computing (SciDAC) program
of the DOE, grant number DE-FC02-01ER41184. A.H. performed this work
under the auspices of the U.S.\ Department of Energy at the Los Alamos
National Laboratory operated by the University of California under
contract No.\ W-7405-ENG-36.  The LANL report number for this article is LA-UR-04-3513.  J.W.M. would like to
thank the Joint Institute for Nuclear Astrophysics (JINA) for providing
a graduate fellowship.


\clearpage
\begin{deluxetable}{cccc}
\tablecaption{Structural Boundaries\label{tab:boundaries}}
\tablewidth{302pt}
\tablehead{\colhead{Mass (M$_\sun$)\tablenotemark{1}} &
\colhead{Fossil Core (M$_\sun$)\tablenotemark{2}} &
\colhead{Fe Core (M$_\sun$)\tablenotemark{3}} &
\colhead{O Base (M$_\sun$)\tablenotemark{4}}}
\startdata
11 & 0.83 & 1.22 (C) & 1.36(C) \\
13 & 0.98 & 1.33 (R) & 1.44(C) \\
15 & 1.05 & 1.41 (R) & 1.74(C) \\
20 & 0.99 & 1.46 (R) & 1.60(C) \\
30 & 1.06 & 1.45 (R) & 2.03(C) \\
35 & 0.96 & 1.47 (M)& 1.66(C) \\
40 & 0.97 & 1.54 (R) & 1.76(C) \\
\enddata
\tablenotetext{1}{The initial mass of the model.}
\tablenotetext{2}{The location of the fossil Fe core, which was formed
  by convective core Si burning, at 5 s before collapse.}
\tablenotetext{3}{The location of the base of the Si burning shell or
  the top of the Fe core, which inludes the fossil Fe core and the
  more recent ashes from the Si burning shell.  A C indicates that the
  shell is convective, R indicates that it is radiative, and M
  inidicates that the shell is radiative in its inner portion and
  convective in its outer portion.}
\tablenotetext{4}{The location of the base of the O burning shell.}

\end{deluxetable}

\clearpage
\begin{deluxetable}{crrrrrr}
\tablecaption{Growing Modes\label{tab:growing}}
\tablewidth{434pt}
\tablehead{\colhead{Mass (M$_\sun$)\tablenotemark{1}} &
\colhead{t (s)\tablenotemark{2}} &
\colhead{Period (s)\tablenotemark{3}} &
\colhead{$\tau$ (s)\tablenotemark{4}} &
\colhead{$1/\tau_{nuc}$ (s$^{-1}$)\tablenotemark{5}} &
\colhead{$-1/\tau_{\nu}$ (s$^{-1}$)\tablenotemark{6}} &
\colhead{$-1/\tau_{leak}$ (s$^{-1}$)\tablenotemark{7}}}
\startdata
\cutinhead{$\ell = 1$\tablenotemark{8}}
11 & 0 &  2.21 & $1.23 \times 10^{1}$ & $1.60 \times 10^{-1}$ & $6.66 \times 10^{-2}$ & $1.23 \times 10^{-2}$ \\
- & 1 &  7.63 & $5.93 \times 10^{2}$ & $6.62 \times 10^{-3}$ & $4.93 \times 10^{-3}$ & $\sim 0$ \\
- & 1 &  4.37 & $1.20 \times 10^{2}$ & $1.52 \times 10^{-2}$ & $6.87 \times 10^{-3}$ & $2.20 \times 10^{-5}$ \\
- & 1 &  3.04 & $5.49 \times 10^{1}$ & $2.87 \times 10^{-2}$ & $6.40 \times 10^{-3}$ & $4.12 \times 10^{-3}$ \\
- & 1 &  2.74 & $1.86 \times 10^{2}$ & $2.56 \times 10^{-2}$ & $6.30 \times 10^{-3}$ & $1.39 \times 10^{-2}$ \\
- & 5 &  4.97 & $1.85 \times 10^{2}$ & $7.09 \times 10^{-3}$ & $1.68 \times 10^{-3}$ & $1.56 \times 10^{-5}$ \\
- & 5 &  3.53 & $2.56 \times 10^{2}$ & $7.96 \times 10^{-3}$ & $2.02 \times 10^{-3}$ & $2.03 \times 10^{-3}$ \\
- & 30 &  9.14 & $3.47 \times 10^{2}$ & $3.79 \times 10^{-3}$ & $9.07 \times 10^{-4}$ & $\sim 0$ \\
- & 30 &  6.24 & $5.69 \times 10^{2}$ & $3.07 \times 10^{-3}$ & $1.31 \times 10^{-3}$ & $\sim 0$ \\
- & 30 &  5.75 & $9.77 \times 10^{1}$ & $1.11 \times 10^{-2}$ & $8.26 \times 10^{-4}$ & $\sim 0$ \\
- & 30 &  4.96 & $8.20 \times 10^{2}$ & $2.68 \times 10^{-3}$ & $1.44 \times 10^{-3}$ & $1.87 \times 10^{-5}$ \\
- & 30 &  4.09 & $1.83 \times 10^{2}$ & $7.23 \times 10^{-3}$ & $9.63 \times 10^{-4}$ & $8.07 \times 10^{-4}$ \\
- & 30 &  3.62 & $1.30 \times 10^{3}$ & $4.14 \times 10^{-3}$ & $1.15 \times 10^{-3}$ & $2.22 \times 10^{-3}$ \\
- & 60 &  9.80 & $6.74 \times 10^{2}$ & $2.09 \times 10^{-3}$ & $6.11 \times 10^{-4}$ & $\sim 0$ \\
- & 60 &  6.80 & $7.24 \times 10^{2}$ & $2.19 \times 10^{-3}$ & $8.11 \times 10^{-4}$ & $\sim 0$ \\
- & 60 &  6.37 & $2.34 \times 10^{2}$ & $4.89 \times 10^{-3}$ & $6.15 \times 10^{-4}$ & $\sim 0$ \\
- & 60 &  4.48 & $8.15 \times 10^{2}$ & $2.36 \times 10^{-3}$ & $6.85 \times 10^{-4}$ & $4.50 \times 10^{-4}$ \\
13 & 0 &  4.44 & $8.59 \times 10^{0}$ & $2.54 \times 10^{-1}$ & $1.37 \times 10^{-1}$ & $6.46 \times 10^{-6}$ \\
- & 0 &  4.34 & $4.70 \times 10^{-1}$ & $2.15 \times 10^{0}$ & $2.32 \times 10^{-2}$ & $9.40 \times 10^{-5}$ \\
- & 1 &  8.42 & $1.55 \times 10^{1}$ & $6.65 \times 10^{-2}$ & $2.19 \times 10^{-3}$ & $4.82 \times 10^{-6}$ \\
- & 1 &  5.59 & $3.47 \times 10^{1}$ & $3.24 \times 10^{-2}$ & $3.38 \times 10^{-3}$ & $1.42 \times 10^{-4}$ \\
- & 5 &  9.50 & $7.90 \times 10^{2}$ & $2.32 \times 10^{-3}$ & $1.05 \times 10^{-3}$ & $4.20 \times 10^{-6}$ \\
15 & 0 &  4.59 & $7.13 \times 10^{-1}$ & $1.43 \times 10^{0}$ & $1.54 \times 10^{-3}$ & $2.61 \times 10^{-2}$ \\
- & 1 &  8.39 & $3.15 \times 10^{1}$ & $3.18 \times 10^{-2}$ & $6.32 \times 10^{-5}$ & $\sim 0$ \\
- & 1 &  5.23 & $5.64 \times 10^{1}$ & $5.02 \times 10^{-2}$ & $6.05 \times 10^{-4}$ & $3.19 \times 10^{-2}$ \\
- & 5 &  8.66 & $1.37 \times 10^{2}$ & $7.35 \times 10^{-3}$ & $4.82 \times 10^{-5}$ & $\sim 0$ \\
- & 30 &  9.15 & $4.43 \times 10^{2}$ & $2.30 \times 10^{-3}$ & $3.88 \times 10^{-5}$ & $\sim 0$ \\
- & 30 &  6.62 & $9.75 \times 10^{2}$ & $1.42 \times 10^{-3}$ & $3.93 \times 10^{-4}$ & $\sim 0$ \\
- & 60 &  9.52 & $6.31 \times 10^{2}$ & $1.63 \times 10^{-3}$ & $4.40 \times 10^{-5}$ & $\sim 0$ \\
- & 60 &  7.03 & $9.44 \times 10^{2}$ & $1.45 \times 10^{-3}$ & $3.92 \times 10^{-4}$ & $\sim 0$ \\
- & 600 & 11.04 & $9.68 \times 10^{5}$ & $6.84 \times 10^{-5}$ & $6.73 \times 10^{-5}$ & $\sim 0$ \\
- & 3600 & 24.28 & $5.48 \times 10^{4}$ & $7.68 \times 10^{-5}$ & $5.86 \times 10^{-5}$ & $\sim 0$ \\
- & 3600 & 16.47 & $9.99 \times 10^{4}$ & $5.89 \times 10^{-5}$ & $4.89 \times 10^{-5}$ & $\sim 0$ \\
- & 3600 & 13.00 & $3.40 \times 10^{5}$ & $2.40 \times 10^{-5}$ & $2.11 \times 10^{-5}$ & $\sim 0$ \\
20 & 1 & 10.03 & $3.61 \times 10^{1}$ & $2.94 \times 10^{-2}$ & $1.65 \times 10^{-3}$ & $1.95 \times 10^{-5}$ \\
- & 1 &  6.77 & $1.77 \times 10^{2}$ & $9.49 \times 10^{-3}$ & $3.68 \times 10^{-3}$ & $1.43 \times 10^{-4}$ \\
- & 5 & 11.14 & $2.91 \times 10^{2}$ & $4.18 \times 10^{-3}$ & $7.44 \times 10^{-4}$ & $5.11 \times 10^{-6}$ \\
25 & 0 &  5.86 & $7.29 \times 10^{-1}$ & $1.38 \times 10^{0}$ & $4.03 \times 10^{-3}$ & $\sim 0$ \\
- & 1 & 10.70 & $2.89 \times 10^{1}$ & $3.79 \times 10^{-2}$ & $3.27 \times 10^{-3}$ & $\sim 0$ \\
- & 1 &  7.09 & $1.74 \times 10^{1}$ & $5.96 \times 10^{-2}$ & $2.20 \times 10^{-3}$ & $\sim 0$ \\
- & 1 &  6.74 & $1.17 \times 10^{2}$ & $1.34 \times 10^{-2}$ & $4.85 \times 10^{-3}$ & $\sim 0$ \\
- & 5 & 11.30 & $3.97 \times 10^{2}$ & $3.91 \times 10^{-3}$ & $1.39 \times 10^{-3}$ & $\sim 0$ \\
- & 5 &  7.88 & $2.54 \times 10^{2}$ & $5.04 \times 10^{-3}$ & $1.10 \times 10^{-3}$ & $\sim 0$ \\
- & 30 &  8.71 & $2.59 \times 10^{3}$ & $1.12 \times 10^{-3}$ & $7.33 \times 10^{-4}$ & $\sim 0$ \\
- & 60 &  9.18 & $1.01 \times 10^{4}$ & $7.30 \times 10^{-4}$ & $6.31 \times 10^{-4}$ & $\sim 0$ \\
- & 3600 & 17.25 & $8.35 \times 10^{5}$ & $6.13 \times 10^{-5}$ & $6.01 \times 10^{-5}$ & $\sim 0$ \\
30 & 0 &  5.35 & $2.28 \times 10^{0}$ & $4.41 \times 10^{-1}$ & $3.31 \times 10^{-3}$ & $\sim 0$ \\
- & 1 &  7.51 & $2.44 \times 10^{1}$ & $4.18 \times 10^{-2}$ & $8.26 \times 10^{-4}$ & $\sim 0$ \\
- & 5 &  8.67 & $4.23 \times 10^{2}$ & $3.04 \times 10^{-3}$ & $6.76 \times 10^{-4}$ & $\sim 0$ \\
- & 30 &  9.79 & $3.84 \times 10^{3}$ & $7.83 \times 10^{-4}$ & $5.23 \times 10^{-4}$ & $\sim 0$ \\
- & 3600 & 16.13 & $5.74 \times 10^{6}$ & $3.21 \times 10^{-5}$ & $3.20 \times 10^{-5}$ & $\sim 0$ \\
35 & 0 &  5.09 & $3.49 \times 10^{1}$ & $9.27 \times 10^{-2}$ & $6.41 \times 10^{-2}$ & $1.49 \times 10^{-5}$ \\
- & 0 &  4.82 & $5.28 \times 10^{-1}$ & $1.90 \times 10^{0}$ & $6.93 \times 10^{-3}$ & $1.08 \times 10^{-3}$ \\
- & 1 &  7.99 & $1.81 \times 10^{2}$ & $8.57 \times 10^{-3}$ & $3.05 \times 10^{-3}$ & $\sim 0$ \\
- & 1 &  7.60 & $1.35 \times 10^{2}$ & $1.06 \times 10^{-2}$ & $3.16 \times 10^{-3}$ & $\sim 0$ \\
- & 1 &  5.85 & $5.22 \times 10^{1}$ & $2.14 \times 10^{-2}$ & $2.20 \times 10^{-3}$ & $\sim 0$ \\
- & 5 &  8.36 & $9.90 \times 10^{2}$ & $2.06 \times 10^{-3}$ & $1.05 \times 10^{-3}$ & $\sim 0$ \\
- & 5 &  6.44 & $1.09 \times 10^{3}$ & $2.14 \times 10^{-3}$ & $1.22 \times 10^{-3}$ & $\sim 0$ \\
- & 30 &  9.19 & $5.03 \times 10^{3}$ & $7.07 \times 10^{-4}$ & $5.09 \times 10^{-4}$ & $\sim 0$ \\
- & 60 &  9.70 & $6.22 \times 10^{3}$ & $5.61 \times 10^{-4}$ & $4.00 \times 10^{-4}$ & $\sim 0$ \\
40 & 0 &  5.95 & $5.45 \times 10^{1}$ & $6.82 \times 10^{-2}$ & $4.98 \times 10^{-2}$ & $1.83 \times 10^{-5}$ \\
- & 0 &  5.47 & $5.43 \times 10^{-1}$ & $1.85 \times 10^{0}$ & $7.20 \times 10^{-3}$ & $1.55 \times 10^{-3}$ \\
- & 1 &  9.00 & $7.88 \times 10^{1}$ & $1.60 \times 10^{-2}$ & $3.31 \times 10^{-3}$ & $\sim 0$ \\
- & 1 &  8.56 & $3.63 \times 10^{1}$ & $3.03 \times 10^{-2}$ & $2.81 \times 10^{-3}$ & $\sim 0$ \\
- & 1 &  7.39 & $4.46 \times 10^{3}$ & $4.74 \times 10^{-3}$ & $4.51 \times 10^{-3}$ & $\sim 0$ \\
- & 1 &  6.36 & $5.41 \times 10^{1}$ & $2.24 \times 10^{-2}$ & $3.47 \times 10^{-3}$ & $5.00 \times 10^{-4}$ \\
- & 1 &  5.95 & $5.62 \times 10^{1}$ & $2.23 \times 10^{-2}$ & $3.36 \times 10^{-3}$ & $1.11 \times 10^{-3}$ \\
- & 5 &  9.45 & $6.06 \times 10^{2}$ & $2.55 \times 10^{-3}$ & $8.97 \times 10^{-4}$ & $\sim 0$ \\
\cutinhead{$\ell = 2$}
11 & 0 &  1.36 & $4.55 \times 10^{1}$ & $1.19 \times 10^{-1}$ & $8.66 \times 10^{-2}$ & $9.96 \times 10^{-3}$ \\
- & 1 &  4.47 & $9.37 \times 10^{2}$ & $6.07 \times 10^{-3}$ & $5.00 \times 10^{-3}$ & $\sim 0$ \\
- & 1 &  2.63 & $1.02 \times 10^{2}$ & $1.57 \times 10^{-2}$ & $5.87 \times 10^{-3}$ & $1.19 \times 10^{-5}$ \\
- & 1 &  1.90 & $1.28 \times 10^{2}$ & $1.75 \times 10^{-2}$ & $7.99 \times 10^{-3}$ & $1.67 \times 10^{-3}$ \\
- & 1 &  1.73 & $7.23 \times 10^{1}$ & $3.58 \times 10^{-2}$ & $4.37 \times 10^{-3}$ & $1.76 \times 10^{-2}$ \\
- & 5 &  4.82 & $2.40 \times 10^{3}$ & $2.28 \times 10^{-3}$ & $1.86 \times 10^{-3}$ & $\sim 0$ \\
- & 5 &  2.98 & $1.98 \times 10^{2}$ & $6.66 \times 10^{-3}$ & $1.59 \times 10^{-3}$ & $1.11 \times 10^{-5}$ \\
- & 5 &  2.68 & $6.02 \times 10^{3}$ & $2.93 \times 10^{-3}$ & $2.74 \times 10^{-3}$ & $2.73 \times 10^{-5}$ \\
- & 5 &  2.21 & $5.67 \times 10^{2}$ & $4.89 \times 10^{-3}$ & $2.40 \times 10^{-3}$ & $7.25 \times 10^{-4}$ \\
- & 5 &  1.98 & $1.41 \times 10^{3}$ & $1.15 \times 10^{-2}$ & $1.31 \times 10^{-3}$ & $9.52 \times 10^{-3}$ \\
- & 30 &  5.35 & $3.45 \times 10^{2}$ & $3.77 \times 10^{-3}$ & $8.73 \times 10^{-4}$ & $\sim 0$ \\
- & 30 &  3.77 & $3.43 \times 10^{3}$ & $1.70 \times 10^{-3}$ & $1.41 \times 10^{-3}$ & $\sim 0$ \\
- & 30 &  3.44 & $8.76 \times 10^{1}$ & $1.21 \times 10^{-2}$ & $7.12 \times 10^{-4}$ & $\sim 0$ \\
- & 30 &  3.07 & $6.34 \times 10^{2}$ & $2.98 \times 10^{-3}$ & $1.39 \times 10^{-3}$ & $1.02 \times 10^{-5}$ \\
- & 30 &  2.55 & $2.50 \times 10^{2}$ & $5.46 \times 10^{-3}$ & $1.11 \times 10^{-3}$ & $3.54 \times 10^{-4}$ \\
- & 30 &  2.29 & $2.94 \times 10^{2}$ & $6.24 \times 10^{-3}$ & $9.68 \times 10^{-4}$ & $1.87 \times 10^{-3}$ \\
- & 60 &  5.74 & $6.82 \times 10^{2}$ & $2.08 \times 10^{-3}$ & $6.13 \times 10^{-4}$ & $\sim 0$ \\
- & 60 &  4.10 & $2.49 \times 10^{3}$ & $1.30 \times 10^{-3}$ & $8.98 \times 10^{-4}$ & $\sim 0$ \\
- & 60 &  3.82 & $1.81 \times 10^{2}$ & $6.04 \times 10^{-3}$ & $5.00 \times 10^{-4}$ & $\sim 0$ \\
- & 60 &  2.78 & $1.21 \times 10^{3}$ & $1.79 \times 10^{-3}$ & $7.59 \times 10^{-4}$ & $2.04 \times 10^{-4}$ \\
- & 60 &  2.50 & $6.38 \times 10^{3}$ & $1.79 \times 10^{-3}$ & $7.31 \times 10^{-4}$ & $8.98 \times 10^{-4}$ \\
13 & 0 &  2.68 & $2.77 \times 10^{0}$ & $4.87 \times 10^{-1}$ & $1.26 \times 10^{-1}$ & $9.02 \times 10^{-6}$ \\
- & 0 &  2.65 & $5.43 \times 10^{-1}$ & $1.88 \times 10^{0}$ & $3.80 \times 10^{-2}$ & $4.10 \times 10^{-5}$ \\
- & 1 &  4.99 & $1.49 \times 10^{1}$ & $6.93 \times 10^{-2}$ & $2.10 \times 10^{-3}$ & $1.89 \times 10^{-6}$ \\
- & 1 &  3.44 & $3.46 \times 10^{1}$ & $3.24 \times 10^{-2}$ & $3.41 \times 10^{-3}$ & $6.10 \times 10^{-5}$ \\
- & 5 &  5.64 & $7.10 \times 10^{2}$ & $2.41 \times 10^{-3}$ & $9.96 \times 10^{-4}$ & $2.04 \times 10^{-6}$ \\
15 & 0 &  2.85 & $6.14 \times 10^{-1}$ & $1.64 \times 10^{0}$ & $1.25 \times 10^{-3}$ & $7.38 \times 10^{-3}$ \\
- & 0 &  1.79 & $1.01 \times 10^{1}$ & $1.76 \times 10^{-1}$ & $6.39 \times 10^{-2}$ & $1.35 \times 10^{-2}$ \\
- & 0 &  1.72 & $3.76 \times 10^{0}$ & $3.35 \times 10^{-1}$ & $3.67 \times 10^{-2}$ & $3.26 \times 10^{-2}$ \\
- & 1 &  6.16 & $5.42 \times 10^{1}$ & $1.85 \times 10^{-2}$ & $3.28 \times 10^{-5}$ & $\sim 0$ \\
- & 1 &  3.24 & $1.89 \times 10^{1}$ & $6.47 \times 10^{-2}$ & $4.80 \times 10^{-4}$ & $1.14 \times 10^{-2}$ \\
- & 5 &  6.34 & $1.75 \times 10^{2}$ & $5.76 \times 10^{-3}$ & $2.55 \times 10^{-5}$ & $\sim 0$ \\
- & 60 &  4.34 & $1.10 \times 10^{3}$ & $1.24 \times 10^{-3}$ & $3.28 \times 10^{-4}$ & $\sim 0$ \\
- & 3600 & 14.17 & $6.48 \times 10^{4}$ & $7.16 \times 10^{-5}$ & $5.62 \times 10^{-5}$ & $\sim 0$ \\
- & 3600 & 10.32 & $1.58 \times 10^{5}$ & $2.71 \times 10^{-5}$ & $2.08 \times 10^{-5}$ & $\sim 0$ \\
- & 3600 &  9.11 & $1.73 \times 10^{5}$ & $4.80 \times 10^{-5}$ & $4.22 \times 10^{-5}$ & $\sim 0$ \\
20 & 1 &  6.04 & $3.50 \times 10^{1}$ & $3.02 \times 10^{-2}$ & $1.65 \times 10^{-3}$ & $9.49 \times 10^{-6}$ \\
- & 1 &  4.19 & $3.43 \times 10^{2}$ & $7.12 \times 10^{-3}$ & $4.15 \times 10^{-3}$ & $4.97 \times 10^{-5}$ \\
25 & 0 &  2.38 & $2.85 \times 10^{0}$ & $4.67 \times 10^{-1}$ & $7.70 \times 10^{-4}$ & $1.16 \times 10^{-1}$ \\
- & 1 &  6.43 & $3.07 \times 10^{1}$ & $3.59 \times 10^{-2}$ & $3.39 \times 10^{-3}$ & $\sim 0$ \\
- & 1 &  4.43 & $1.52 \times 10^{1}$ & $6.78 \times 10^{-2}$ & $1.87 \times 10^{-3}$ & $\sim 0$ \\
- & 1 &  4.24 & $2.20 \times 10^{2}$ & $9.57 \times 10^{-3}$ & $5.02 \times 10^{-3}$ & $\sim 0$ \\
- & 5 &  6.58 & $1.21 \times 10^{3}$ & $2.49 \times 10^{-3}$ & $1.67 \times 10^{-3}$ & $\sim 0$ \\
- & 5 &  4.91 & $2.32 \times 10^{2}$ & $5.39 \times 10^{-3}$ & $1.07 \times 10^{-3}$ & $\sim 0$ \\
- & 30 &  5.41 & $1.56 \times 10^{3}$ & $1.17 \times 10^{-3}$ & $5.28 \times 10^{-4}$ & $\sim 0$ \\
- & 60 &  5.70 & $8.32 \times 10^{3}$ & $7.60 \times 10^{-4}$ & $6.40 \times 10^{-4}$ & $\sim 0$ \\
- & 3600 &  4.59 & $4.81 \times 10^{4}$ & $3.23 \times 10^{-5}$ & $1.15 \times 10^{-5}$ & $\sim 0$ \\
30 & 0 &  3.20 & $2.27 \times 10^{0}$ & $4.41 \times 10^{-1}$ & $1.12 \times 10^{-3}$ & $\sim 0$ \\
- & 0 &  1.68 & $1.49 \times 10^{0}$ & $7.94 \times 10^{-1}$ & $2.55 \times 10^{-3}$ & $1.21 \times 10^{-1}$ \\
- & 1 &  4.52 & $2.47 \times 10^{1}$ & $4.12 \times 10^{-2}$ & $6.75 \times 10^{-4}$ & $\sim 0$ \\
- & 5 &  5.23 & $4.11 \times 10^{2}$ & $3.01 \times 10^{-3}$ & $5.74 \times 10^{-4}$ & $\sim 0$ \\
- & 30 &  5.91 & $3.37 \times 10^{3}$ & $7.55 \times 10^{-4}$ & $4.58 \times 10^{-4}$ & $\sim 0$ \\
35 & 0 &  2.98 & $4.93 \times 10^{-1}$ & $2.03 \times 10^{0}$ & $5.28 \times 10^{-3}$ & $6.57 \times 10^{-4}$ \\
- & 1 &  4.77 & $2.31 \times 10^{2}$ & $7.53 \times 10^{-3}$ & $3.20 \times 10^{-3}$ & $\sim 0$ \\
- & 1 &  4.56 & $1.34 \times 10^{2}$ & $1.06 \times 10^{-2}$ & $3.15 \times 10^{-3}$ & $\sim 0$ \\
- & 1 &  3.63 & $5.13 \times 10^{1}$ & $2.17 \times 10^{-2}$ & $2.23 \times 10^{-3}$ & $\sim 0$ \\
- & 1 &  3.41 & $2.01 \times 10^{3}$ & $4.80 \times 10^{-3}$ & $4.30 \times 10^{-3}$ & $\sim 0$ \\
- & 5 &  5.00 & $8.02 \times 10^{2}$ & $2.24 \times 10^{-3}$ & $9.89 \times 10^{-4}$ & $\sim 0$ \\
- & 5 &  3.99 & $1.45 \times 10^{3}$ & $1.96 \times 10^{-3}$ & $1.27 \times 10^{-3}$ & $\sim 0$ \\
- & 30 &  5.49 & $4.08 \times 10^{3}$ & $7.37 \times 10^{-4}$ & $4.92 \times 10^{-4}$ & $\sim 0$ \\
- & 60 &  5.79 & $4.78 \times 10^{3}$ & $5.93 \times 10^{-4}$ & $3.84 \times 10^{-4}$ & $\sim 0$ \\
40 & 0 &  3.44 & $5.12 \times 10^{-1}$ & $1.96 \times 10^{0}$ & $6.00 \times 10^{-3}$ & $7.37 \times 10^{-4}$ \\
- & 0 &  2.04 & $1.14 \times 10^{1}$ & $2.73 \times 10^{-1}$ & $1.27 \times 10^{-3}$ & $1.83 \times 10^{-1}$ \\
- & 1 &  5.37 & $7.42 \times 10^{1}$ & $1.68 \times 10^{-2}$ & $3.32 \times 10^{-3}$ & $\sim 0$ \\
- & 1 &  5.14 & $3.53 \times 10^{1}$ & $3.11 \times 10^{-2}$ & $2.85 \times 10^{-3}$ & $\sim 0$ \\
- & 1 &  3.99 & $6.50 \times 10^{1}$ & $1.94 \times 10^{-2}$ & $3.74 \times 10^{-3}$ & $2.70 \times 10^{-4}$ \\
- & 1 &  3.75 & $4.21 \times 10^{1}$ & $2.75 \times 10^{-2}$ & $2.95 \times 10^{-3}$ & $8.16 \times 10^{-4}$ \\
- & 5 &  5.68 & $5.29 \times 10^{2}$ & $2.75 \times 10^{-3}$ & $8.57 \times 10^{-4}$ & $\sim 0$ \\
\cutinhead{$\ell = 3$}
11 & 0 &  0.99 & $1.10 \times 10^{1}$ & $1.63 \times 10^{-1}$ & $5.81 \times 10^{-2}$ & $1.35 \times 10^{-2}$ \\
- & 0 &  0.88 & $3.74 \times 10^{0}$ & $4.71 \times 10^{-1}$ & $1.03 \times 10^{-2}$ & $1.94 \times 10^{-1}$ \\
- & 1 &  1.95 & $1.12 \times 10^{2}$ & $1.42 \times 10^{-2}$ & $5.20 \times 10^{-3}$ & $2.74 \times 10^{-6}$ \\
- & 1 &  1.33 & $2.69 \times 10^{1}$ & $4.67 \times 10^{-2}$ & $3.09 \times 10^{-3}$ & $6.43 \times 10^{-3}$ \\
- & 5 &  3.49 & $1.49 \times 10^{3}$ & $2.34 \times 10^{-3}$ & $1.66 \times 10^{-3}$ & $\sim 0$ \\
- & 5 &  2.21 & $2.90 \times 10^{2}$ & $5.19 \times 10^{-3}$ & $1.73 \times 10^{-3}$ & $1.77 \times 10^{-6}$ \\
- & 5 &  2.04 & $7.05 \times 10^{2}$ & $3.85 \times 10^{-3}$ & $2.43 \times 10^{-3}$ & $6.93 \times 10^{-6}$ \\
- & 5 &  1.72 & $3.69 \times 10^{3}$ & $2.98 \times 10^{-3}$ & $2.63 \times 10^{-3}$ & $8.16 \times 10^{-5}$ \\
- & 5 &  1.53 & $9.65 \times 10^{1}$ & $1.46 \times 10^{-2}$ & $1.06 \times 10^{-3}$ & $3.19 \times 10^{-3}$ \\
- & 30 &  3.86 & $3.89 \times 10^{2}$ & $3.45 \times 10^{-3}$ & $8.84 \times 10^{-4}$ & $\sim 0$ \\
- & 30 &  2.54 & $8.66 \times 10^{1}$ & $1.22 \times 10^{-2}$ & $6.88 \times 10^{-4}$ & $\sim 0$ \\
- & 30 &  2.34 & $3.47 \times 10^{2}$ & $4.18 \times 10^{-3}$ & $1.30 \times 10^{-3}$ & $\sim 0$ \\
- & 30 &  1.98 & $3.78 \times 10^{2}$ & $3.92 \times 10^{-3}$ & $1.22 \times 10^{-3}$ & $4.68 \times 10^{-5}$ \\
- & 30 &  1.79 & $1.43 \times 10^{2}$ & $8.31 \times 10^{-3}$ & $8.23 \times 10^{-4}$ & $4.93 \times 10^{-4}$ \\
- & 60 &  4.15 & $8.29 \times 10^{2}$ & $1.86 \times 10^{-3}$ & $6.54 \times 10^{-4}$ & $\sim 0$ \\
- & 60 &  2.83 & $1.53 \times 10^{2}$ & $6.97 \times 10^{-3}$ & $4.27 \times 10^{-4}$ & $\sim 0$ \\
- & 60 &  2.56 & $5.36 \times 10^{4}$ & $9.63 \times 10^{-4}$ & $9.45 \times 10^{-4}$ & $\sim 0$ \\
- & 60 &  2.16 & $2.59 \times 10^{3}$ & $1.23 \times 10^{-3}$ & $8.23 \times 10^{-4}$ & $2.29 \times 10^{-5}$ \\
- & 60 &  1.95 & $7.26 \times 10^{2}$ & $2.24 \times 10^{-3}$ & $6.45 \times 10^{-4}$ & $2.18 \times 10^{-4}$ \\
13 & 0 &  2.02 & $9.43 \times 10^{-1}$ & $1.14 \times 10^{0}$ & $8.20 \times 10^{-2}$ & $3.31 \times 10^{-6}$ \\
- & 0 &  2.01 & $9.39 \times 10^{-1}$ & $1.15 \times 10^{0}$ & $8.29 \times 10^{-2}$ & $3.63 \times 10^{-6}$ \\
- & 0 &  2.01 & $9.39 \times 10^{-1}$ & $1.15 \times 10^{0}$ & $8.29 \times 10^{-2}$ & $3.63 \times 10^{-6}$ \\
- & 1 &  3.66 & $1.41 \times 10^{1}$ & $7.29 \times 10^{-2}$ & $1.99 \times 10^{-3}$ & $3.30 \times 10^{-7}$ \\
- & 1 &  2.65 & $3.66 \times 10^{1}$ & $3.09 \times 10^{-2}$ & $3.53 \times 10^{-3}$ & $7.66 \times 10^{-6}$ \\
- & 5 &  4.14 & $6.59 \times 10^{2}$ & $2.47 \times 10^{-3}$ & $9.52 \times 10^{-4}$ & $3.54 \times 10^{-7}$ \\
15 & 0 &  2.21 & $5.65 \times 10^{-1}$ & $1.77 \times 10^{0}$ & $1.06 \times 10^{-3}$ & $1.26 \times 10^{-3}$ \\
- & 0 &  1.42 & $3.08 \times 10^{1}$ & $1.02 \times 10^{-1}$ & $6.84 \times 10^{-2}$ & $1.45 \times 10^{-3}$ \\
- & 0 &  1.38 & $4.50 \times 10^{0}$ & $2.57 \times 10^{-1}$ & $3.07 \times 10^{-2}$ & $4.45 \times 10^{-3}$ \\
- & 1 &  2.49 & $1.31 \times 10^{1}$ & $7.92 \times 10^{-2}$ & $3.38 \times 10^{-4}$ & $2.56 \times 10^{-3}$ \\
- & 1 &  1.67 & $2.17 \times 10^{1}$ & $5.90 \times 10^{-2}$ & $1.34 \times 10^{-3}$ & $1.16 \times 10^{-2}$ \\
- & 5 &  2.76 & $1.89 \times 10^{2}$ & $7.16 \times 10^{-3}$ & $2.20 \times 10^{-4}$ & $1.64 \times 10^{-3}$ \\
- & 30 &  3.06 & $1.05 \times 10^{3}$ & $1.22 \times 10^{-3}$ & $2.67 \times 10^{-4}$ & $\sim 0$ \\
- & 60 &  3.26 & $1.07 \times 10^{3}$ & $1.20 \times 10^{-3}$ & $2.67 \times 10^{-4}$ & $\sim 0$ \\
- & 3600 & 10.18 & $8.41 \times 10^{4}$ & $6.51 \times 10^{-5}$ & $5.32 \times 10^{-5}$ & $\sim 0$ \\
- & 3600 &  8.42 & $2.73 \times 10^{5}$ & $8.11 \times 10^{-6}$ & $4.45 \times 10^{-6}$ & $\sim 0$ \\
- & 3600 &  6.95 & $5.68 \times 10^{5}$ & $5.03 \times 10^{-5}$ & $4.85 \times 10^{-5}$ & $\sim 0$ \\
20 & 1 &  4.58 & $6.79 \times 10^{1}$ & $1.79 \times 10^{-2}$ & $3.14 \times 10^{-3}$ & $7.57 \times 10^{-7}$ \\
- & 1 &  4.51 & $6.96 \times 10^{1}$ & $1.76 \times 10^{-2}$ & $3.25 \times 10^{-3}$ & $8.71 \times 10^{-7}$ \\
- & 1 &  3.35 & $2.38 \times 10^{2}$ & $8.13 \times 10^{-3}$ & $3.91 \times 10^{-3}$ & $5.18 \times 10^{-6}$ \\
25 & 0 &  2.00 & $2.80 \times 10^{0}$ & $3.82 \times 10^{-1}$ & $3.23 \times 10^{-4}$ & $2.42 \times 10^{-2}$ \\
- & 0 &  1.00 & $1.17 \times 10^{2}$ & $1.99 \times 10^{-2}$ & $9.07 \times 10^{-3}$ & $2.28 \times 10^{-3}$ \\
- & 1 &  3.45 & $1.21 \times 10^{1}$ & $8.42 \times 10^{-2}$ & $1.48 \times 10^{-3}$ & $\sim 0$ \\
- & 1 &  2.20 & $2.17 \times 10^{2}$ & $2.06 \times 10^{-2}$ & $2.23 \times 10^{-3}$ & $1.37 \times 10^{-2}$ \\
- & 1 &  2.17 & $3.40 \times 10^{2}$ & $1.74 \times 10^{-2}$ & $2.94 \times 10^{-3}$ & $1.15 \times 10^{-2}$ \\
- & 5 &  3.81 & $1.92 \times 10^{2}$ & $6.24 \times 10^{-3}$ & $1.03 \times 10^{-3}$ & $\sim 0$ \\
- & 30 &  4.20 & $2.15 \times 10^{3}$ & $1.24 \times 10^{-3}$ & $7.70 \times 10^{-4}$ & $\sim 0$ \\
- & 60 &  4.41 & $9.52 \times 10^{3}$ & $7.75 \times 10^{-4}$ & $6.70 \times 10^{-4}$ & $\sim 0$ \\
- & 600 &  3.18 & $2.42 \times 10^{3}$ & $4.36 \times 10^{-4}$ & $2.25 \times 10^{-5}$ & $\sim 0$ \\
- & 3600 &  3.86 & $4.35 \times 10^{4}$ & $3.23 \times 10^{-5}$ & $9.29 \times 10^{-6}$ & $\sim 0$ \\
30 & 0 &  2.36 & $2.30 \times 10^{0}$ & $4.35 \times 10^{-1}$ & $6.71 \times 10^{-4}$ & $\sim 0$ \\
- & 0 &  1.38 & $1.10 \times 10^{0}$ & $9.31 \times 10^{-1}$ & $1.52 \times 10^{-3}$ & $2.41 \times 10^{-2}$ \\
- & 1 &  3.34 & $2.44 \times 10^{1}$ & $4.16 \times 10^{-2}$ & $5.30 \times 10^{-4}$ & $\sim 0$ \\
- & 1 &  1.66 & $6.45 \times 10^{1}$ & $4.43 \times 10^{-2}$ & $4.29 \times 10^{-4}$ & $2.83 \times 10^{-2}$ \\
- & 5 &  3.87 & $3.90 \times 10^{2}$ & $3.01 \times 10^{-3}$ & $4.51 \times 10^{-4}$ & $\sim 0$ \\
- & 30 &  4.38 & $2.59 \times 10^{3}$ & $7.54 \times 10^{-4}$ & $3.68 \times 10^{-4}$ & $\sim 0$ \\
35 & 0 &  2.30 & $4.93 \times 10^{-1}$ & $2.03 \times 10^{0}$ & $2.93 \times 10^{-3}$ & $8.37 \times 10^{-5}$ \\
- & 1 &  3.55 & $6.73 \times 10^{3}$ & $4.07 \times 10^{-3}$ & $3.92 \times 10^{-3}$ & $\sim 0$ \\
- & 1 &  3.41 & $1.11 \times 10^{2}$ & $1.20 \times 10^{-2}$ & $2.98 \times 10^{-3}$ & $\sim 0$ \\
- & 1 &  2.79 & $4.35 \times 10^{1}$ & $2.49 \times 10^{-2}$ & $1.93 \times 10^{-3}$ & $\sim 0$ \\
- & 1 &  2.65 & $2.13 \times 10^{4}$ & $4.25 \times 10^{-3}$ & $4.20 \times 10^{-3}$ & $\sim 0$ \\
- & 5 &  3.71 & $7.14 \times 10^{2}$ & $2.34 \times 10^{-3}$ & $9.42 \times 10^{-4}$ & $\sim 0$ \\
- & 5 &  3.07 & $1.42 \times 10^{3}$ & $1.94 \times 10^{-3}$ & $1.23 \times 10^{-3}$ & $\sim 0$ \\
- & 30 &  4.06 & $6.22 \times 10^{3}$ & $6.79 \times 10^{-4}$ & $5.18 \times 10^{-4}$ & $\sim 0$ \\
- & 60 &  4.28 & $6.63 \times 10^{3}$ & $5.36 \times 10^{-4}$ & $3.85 \times 10^{-4}$ & $\sim 0$ \\
- & 3600 &  3.31 & $1.63 \times 10^{5}$ & $1.14 \times 10^{-5}$ & $5.26 \times 10^{-6}$ & $\sim 0$ \\
40 & 0 &  1.74 & $4.63 \times 10^{0}$ & $2.59 \times 10^{-1}$ & $3.73 \times 10^{-4}$ & $4.29 \times 10^{-2}$ \\
- & 0 &  0.98 & $2.71 \times 10^{2}$ & $2.07 \times 10^{-2}$ & $1.51 \times 10^{-2}$ & $1.89 \times 10^{-3}$ \\
- & 1 &  4.00 & $1.24 \times 10^{2}$ & $1.20 \times 10^{-2}$ & $3.90 \times 10^{-3}$ & $\sim 0$ \\
- & 1 &  3.86 & $2.85 \times 10^{1}$ & $3.77 \times 10^{-2}$ & $2.67 \times 10^{-3}$ & $\sim 0$ \\
- & 1 &  3.55 & $2.18 \times 10^{3}$ & $5.41 \times 10^{-3}$ & $4.95 \times 10^{-3}$ & $5.42 \times 10^{-7}$ \\
- & 1 &  3.09 & $4.21 \times 10^{1}$ & $2.70 \times 10^{-2}$ & $3.22 \times 10^{-3}$ & $6.66 \times 10^{-5}$ \\
- & 1 &  2.94 & $4.82 \times 10^{1}$ & $2.41 \times 10^{-2}$ & $3.23 \times 10^{-3}$ & $1.16 \times 10^{-4}$ \\
- & 5 &  4.25 & $4.14 \times 10^{2}$ & $3.18 \times 10^{-3}$ & $7.66 \times 10^{-4}$ & $\sim 0$ \\
- & 5 &  3.20 & $3.03 \times 10^{3}$ & $1.66 \times 10^{-3}$ & $1.22 \times 10^{-3}$ & $1.08 \times 10^{-4}$ \\
- & 30 &  4.65 & $1.90 \times 10^{4}$ & $4.75 \times 10^{-4}$ & $4.23 \times 10^{-4}$ & $\sim 0$ \\
\enddata
\tablenotetext{1}{The initial mass of the model.}
\tablenotetext{2}{The time before collapse in seconds.}
\tablenotetext{3}{The period of the particular growing mode.}
\tablenotetext{4}{The growth timescale given by $1/\tau = 1/\tau_{nuc}
+ 1/\tau_{\nu} + 1/\tau_{leak}$}
\tablenotetext{5}{Inverse of the growth timescale from nuclear
burning.}
\tablenotetext{6}{Negative and inverse of the decay timescale due to
$\nu$ emission.}
\tablenotetext{7}{Negative and inverse of the decay timescale due to
acoustic losses to the envelope.}
\tablenotetext{8}{Each row is associated with a specific $\ell$ given
by the header for each section.}

\end{deluxetable}

\clearpage

\end{document}